\pdfoutput=1
\documentclass[aps,prd,amsmath,floats,floatfix,onecolumn,superscriptaddress,nofootinbib,showpacs]{revtex4-2}

\usepackage{comment}
\usepackage[T1]{fontenc}
\usepackage[utf8]{inputenc}
\usepackage{lmodern}
\usepackage{lipsum}
\usepackage{physics}
\usepackage[dvipsnames, usenames]{xcolor}
\definecolor{linkcolor}{rgb}{0.0,0.3,0.5}
\usepackage[hypertexnames=false, unicode, colorlinks=true, linkcolor=linkcolor,
citecolor=linkcolor, filecolor=linkcolor,urlcolor=linkcolor,
pdfusetitle]{hyperref}

\usepackage[all]{hypcap}
\usepackage{graphicx}
\usepackage{xspace}
\usepackage{amssymb}
\usepackage[normalem]{ulem} 
\usepackage{bm} 
\usepackage{appendix}
\usepackage[small]{caption}
\usepackage{subcaption}

\usepackage{microtype}

\usepackage[english]{babel}
\usepackage{blindtext}

\graphicspath{%
  {figs/}%
}

\usepackage{tikz}
\usetikzlibrary{shapes,snakes}
\usetikzlibrary{arrows,intersections,patterns,decorations.markings,shapes.geometric}
\usetikzlibrary{calc,fadings,decorations.pathreplacing,positioning}
\usetikzlibrary{decorations.pathmorphing}
\tikzset{snake it/.style={decorate, decoration=snake}}

\usepackage{pgfplots}
\usepgfplotslibrary{colormaps}
\usetikzlibrary{intersections,patterns,pgfplots.fillbetween}

\tikzset{->-/.style={decoration={
  markings,
  mark=at position .5 with {\arrow{>}}},postaction={decorate}}
}
\tikzset{-<-/.style={decoration={
  markings,
  mark=at position .5 with {\arrow{<}}},postaction={decorate}}
}

\tikzset{%
  >=latex, 
  inner sep=0pt,%
  outer sep=2pt,%
  mark coordinate/.style={inner sep=0pt,outer sep=0pt,minimum size=3pt,
    fill=black,circle}%
}

\DeclareMathAlphabet{\mathpzc}{OT1}{pzc}{m}{it}

\definecolor{darkred}{RGB}{175,0,0}
\definecolor{darkblue}{RGB}{14,0,185}
\definecolor{salmon}{RGB}{255,160,105}

\definecolor{darkblue}{RGB}{14,0,185}

\newcommand*{\bk}{\mathbf{k}}
\newcommand*{\bq}{\mathbf{q}}
\newcommand*{\bp}{\mathbf{p}}

\newcommand*{\bx}{\mathbf{x}}

\newcommand{\beq}{\begin{equation}}
\newcommand{\eeq}{\end{equation}}

\def\k{{\boldsymbol{k}}}
\def\q{{\boldsymbol{q}}}
\def\p{{\boldsymbol{p}}}
\def\r{{\boldsymbol{r}}}

\def\x{{\boldsymbol{x}}}
\def\y{{\boldsymbol{y}}}

\begin{document}

\title{Equivalence of the field-level inference and conventional analyses on large scales}

\author{Francesco Spezzati}
\affiliation{Dipartimento di Fisica Galileo Galilei, Universit\` a di Padova, I-35131 Padova, Italy}
\affiliation{INFN Sezione di Padova, I-35131 Padova, Italy}
\author{Marco Marinucci}
\affiliation{Dipartimento di Fisica Galileo Galilei, Universit\` a di Padova, I-35131 Padova, Italy}
\affiliation{INFN Sezione di Padova, I-35131 Padova, Italy}
\author{Marko Simonović}
\affiliation{Dipartimento di Fisica e Astronomia, Universit\`a di Firenze, I-50019 Sesto Fiorentino, Italy}
\affiliation{INFN, Sezione di Firenze, I-50019 Sesto Fiorentino, Italy}

\begin{abstract}
We study a simple setup with dark matter halos in real space, with the amplitude of the linear density field~$A$ as the only free cosmological parameter. We show that Eulerian perturbation theory is adequate for describing this system on large scales, compute the leading~$n$-point functions and perform a joint power spectrum, bispectrum and trispectrum analysis. Beyond the bispectrum which is crucial for breaking the degeneracy between~$A$ and the linear bias, we find that addition of the trispectrum reduces the error on~$A$ by only~$20-30\%$. Our results for the joint analysis are in good agreement with recent field-level analyses in the same setup. This implies that the field-level inference on large scales does not get significant information from large displacements beyond those in Eulerian kernels or higher-order~$n$-point functions beyond the trispectrum. We provide further evidence for this showing that the dependence of the error bars on the maximum wavenumbers used in the analysis is the same in the two approaches. Our results are in disagreement with some of the recent joint power spectrum and bispectrum analyses using likelihood-free inference based on perturbative forward modeling. We discuss a possible origin of this discrepancy and highlight the importance of resolving it in order to have the optimal results in cosmological analyses based on perturbation theory. 
\end{abstract}

\maketitle

\section{Introduction and main results}
\noindent Field-level Bayesian inference (FBI) of cosmological parameters has recently emerged as an alternative to conventional $n$-point function analyses. This approach offers several advantages, such as being optimal and allowing for a simpler inclusion of data systematics. However, in the context of galaxy clustering, field-level inference which relies on detailed cosmological simulations of the observed galaxy density field remains infeasible, due to very high computational cost. A practical alternative is FBI based on perturbative forward modeling. Within the framework of the Effective Field Theory of Large-Scale Structure (EFTofLSS)~\cite{Baumann:2010tm,Carrasco:2012cv,Assassi:2014fva,Senatore:2014eva,Mirbabayi:2014zca,Senatore:2014vja,Manzotti:2014loa,Ivanov:2022mrd,Cabass:2022avo}, perturbative forward modeling~\cite{Baldauf:2015tla,Baldauf:2015zga,Schmittfull:2018yuk,Schmidt:2018bkr,Modi:2019hnu,Schmittfull:2020trd,Schmidt:2020ovm,Obuljen:2022cjo,Stadler:2024fui,Stadler:2024aff} provides a universal and reliable description of the galaxy density field on large scales with a handful of free parameters to be fitted from the data. Considerable effort has been made to test FBI with perturbative forward modeling and apply it in practice~\cite{Cabass:2019lqx,Elsner:2019rql,Schmidt:2020viy,Cabass:2020nwf,Schmidt:2020tao,Kostic:2022vok,Babic:2022dws,Stadler:2023hea,Tucci:2023bag,Babic:2024wph} (see also~\cite{Peron:2025lgh} for a recent application of the LSS bootstrap~\cite{DAmico:2021rdb,Amendola:2023awr,Marinucci:2024add,Ansari:2025nsf} to FBI). 

A notable recent cosmological analysis using FBI was presented in~\cite{Nguyen:2024yth}, focusing on comparison of the field-level and conventional~$n$-point function based approaches. By using the exact same perturbative forward model in both FBI and the simulation-based\footnote{Here ``simulation'' refers not to the real cosmological simulations, but generation of nonlinear realizations based on the same perturbative forward model as in FBI. For more details, see~\cite{Nguyen:2024yth}.} inference (SBI) for the power spectrum (P) and bispectrum (B), the authors found that FBI achieves a 3-5 times more precise measurement of the amplitude of the initial density field~$A$ than the standard P+B combination. The leading hypotheses for explaining this dramatic difference is that the FBI uses more information from higher order $n$-point functions and exploits at the field level some non-perturbative features of the nonlinear evolution which come with no new free parameters (e.g.,~large displacements). However, due to complexity of both FBI and SBI even in the perturbative framework, this has not been rigorously proven and a clear understanding of where would the additional information in FBI exactly come from is still lacking. 

Similar analysis was presented in the data challenge for beyond the two-point galaxy clustering statistics~\cite{Beyond-2pt:2024mqz}. For example, Fig.~22 in~\cite{Beyond-2pt:2024mqz} shows the constraints on the amplitude of the linear field for the FBI and the standard EFT-based joint power spectrum and bispectrum analysis for the same volume as in~\cite{Nguyen:2024yth}, although with less dense sample. In this case the difference between the two methods is less dramatic (see the Sec.~6.3 of~\cite{Beyond-2pt:2024mqz} for more details about the comparison), but it is hard to straightforwardly compare them. This is mainly due to fact that the $k_{\rm max}$ used in the P+B analysis is larger from the scale cuts in the FBI. In addition, for this larger~$k_{\rm max}$ in the ``restricted'' EFT P+B analysis only the errors are presented (the posterior is centered at the best-fit values for the analysis with lower~$k_{\rm max}$). This leaves several open questions. Is the P+B analysis biased for smaller~$k_{\rm max}$? Are the error bars different for smaller~$k_{\rm max}$? And how do they compare to the FBI results and the SBI errors from~\cite{Nguyen:2024yth}? Providing clear answers is very important, both to get a better understanding of theoretical aspects of FBI and to get firmer basis for its applications in practice and comparison to the conventional analyses.   

The goal of this paper is to shed new light on these questions, focusing on the usual perturbation theory P+B analysis in the extremely simplified but realistic settings. We use a simple, tractable and intuitive Eulerian perturbative model (without infrared resummation), with the simple Gaussian likelihood and covariance. Validity of such model and likelihood on large scales can be easily verified in practice and we will provide several arguments for it. In a nutshell, such simplified approach is possible given that the analysis presented in~\cite{Nguyen:2024yth} follows a particularly simple setup. It uses a simulation box with a relatively small volume of $V=(2\;{\rm Gpc}/h)^3$, which implies relatively large statistical error bars. The studied sample consists of dark matter halos in real space with a very high number density. Additionally, it focuses only on relatively high redshifts $z\geq 0.5$ and large scales, with the maximum wavenumber used in the analysis being $k_{\rm max}\leq 0.12 \;h/{\rm Mpc}$. 
Such setup with regular periodic box, low shot noise, simple tracers without redshift-space distortions or other observational effects and small fluctuations on very large scales, is ideal for understanding FBI using simple analytical methods. Importantly, some of our results then can be easily derived in a very clear and transparent way, without the need to relay on MCMC analyses only. Our approach and results crucially rely on several key assumptions that simplify the description of the system significantly. We would like to clearly state them at the very beginning, given that we will use them throughout the paper.  

First, we assume that the field of dark matter halos in real space for given data cuts and redshifts is well described by Eulerian perturbation theory. Although the forward model used in~\cite{Nguyen:2024yth} is based on Lagrangian perturbation theory, one can show that the two schemes are practically indistinguishable on large scales, given the statistical error bars of the specified simulation box with $V=(2\;{\rm Gpc}/h)^3$. We provide justification and more details in Sec.~\ref{sec:EPTvsLPT}. An important consequence of this fact is that the large displacements beyond those contained in Eulerian kernels cannot play a major role in explaining the results of~\cite{Nguyen:2024yth}.

Assuming that the Eulerian description is sufficient, we further assume that all nonlinearities are controlled by only one parameter\footnote{Other parameters, such as the Lagrangian radius of dark matter halos, are irrelevant, given the large number density of the sample, which implies low halo mass.}---the variance of the density field~$\Delta^2$ at the scale~$k$---and that this parameter is small. It is given by
\begin{equation}
\Delta^2(k) = \frac{1}{2\pi^2} \int_0^{k} q^2 dq P_{\rm lin}(q) \;,
\end{equation}
where~$P_{\rm lin}(k)$ is the linear matter power spectrum. In a $\Lambda$CDM-like cosmology, for $k_{\rm max}=0.12\;h/{\rm Mpc}$ and~$z=0.5$, the variance of the density field is approximately $\Delta^2\approx 0.13$, and it gets smaller at higher redshifts and for smaller values of~$k_{\rm max}$. Importantly, we do not only assume that the relative size of each individual term in perturbative calculation is given by~$\Delta^2(k_{\rm max})$, but that the {\em total} loop contribution to the correlation functions has the same size. This is a nontrivial assumption, since the total one-loop contributions to the higher order $n$-point function can have a large number of terms multiplied by a large number of different nuisance parameters. In other words, we assume that while all bias parameters and counterterms can be numbers of order~$\mathcal{O}(1)$, for a realistic tracers they cannot conspire to give large loop corrections. We will explicitly check that this is the case for a typical sample using large-volume numerical simulations and show that deviations of measured correlation functions from the leading order perturbative predictions are roughly given by~$\Delta^2(k)$ on large scales. Even though our assumption puts some constraints on the allowed values of nuisance parameters, it is essential for the perturbation theory to make sense and it is sometimes referred to as the perturbativity prior~\cite{Braganca:2023pcp}. 

\begin{figure}
    \centering
    \includegraphics[width=0.5\linewidth]{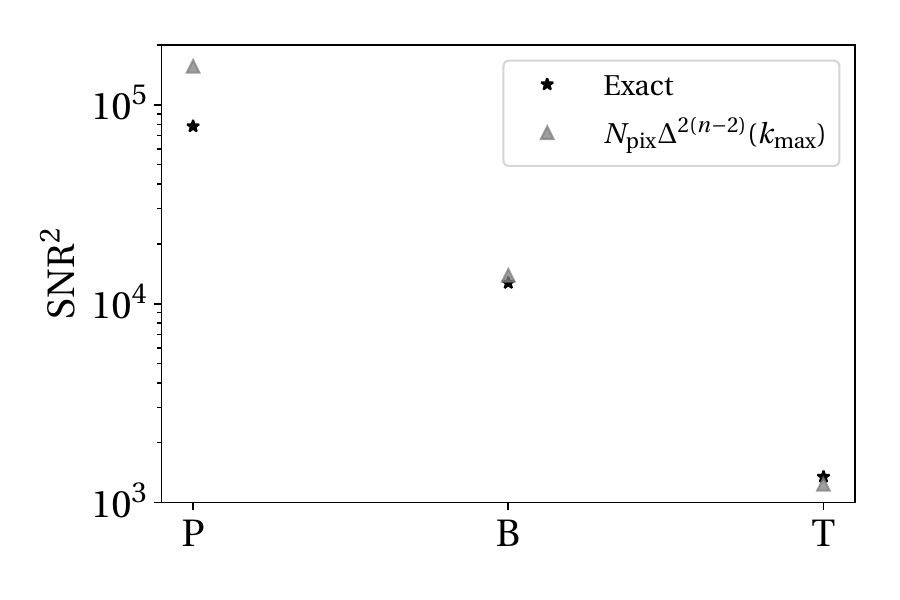}
    \caption{SNR for the power spectrum, the bispectrum and the trispectrum. The stars correspond to the exact evaluation using the standard Gaussian likelihood described in Sec.~\ref{sec:likelihood} (using fiducial bias parameters from Sec.~\ref{sec:results}), while the triangles correspond to the estimate in Eq.~\eqref{eq:snrest}. The agreement between the two for the higher-order~$n$-point functions is quite good. A factor of 2 difference for the power spectrum is expected, since the exact result for the SNR is given by~$({\rm SNR})_2^2= N_{\rm pix.}/2$. }
    \label{fig:SNR}
\end{figure}

With the Eulerian forward model where all nonlinearities are controlled by a single parameter $\Delta^2$ it is easy to compute a few leading order correlation functions. Since we are interested in comparison to the FBI analysis of~\cite{Nguyen:2024yth}, we use the cubic model. Within this model, one can consistently compute the one-loop power spectrum, the tree-level bispectrum and the tree-level trispectrum (T). Higher-order $n$-point functions and additional loop corrections require quartic and higher-order
nonlinearities. Furthermore, since we are focused on large scales only, their impact on cosmological constraints should be much less significant. To estimate this, we can approximate the signal-to-noise ratio (SNR) in a connected $n$-point function of the galaxy density field~$\delta_g$ as follows
\begin{equation}\label{eq:snrest}
({\rm SNR})^2_n \approx \int^{k_{\rm max}} \frac{d^3 \k_1}{(2\pi)^3} \cdots \int^{k_{\rm max}} \frac{d^3 \k_n}{(2\pi)^3} \frac{\langle \delta_g(\k_1) \cdots \delta_g(\k_n) \rangle_c^2}{P_g(k_1) \cdots P_g(k_n)} \sim N_{\rm pix.} \Delta^{2(n-2)}(k_{\rm max})\big[ 1+ \mathcal O (\Delta^2(k_{\rm max})) \big] \;, 
\end{equation}
where $P_g(k)$ is the galaxy power spectrum, $N_{\rm pix.} \equiv V \int^{k_{\rm max}} d^3 \k/(2\pi)^3$ is the number of pixels or Fourier modes in a survey and $\Delta^2(k_{\rm max})\lesssim0.1$ for the range of redshifts and scales of interest. Note that this estimate is correct up to~$\mathcal{O}(1)$ overall multiplicative factor which depends on fiducial values of bias parameters which we neglect for simplicity. For example, for the power spectrum~($n=2$), the correct estimate is~$({\rm SNR})_2^2= N_{\rm pix.}/2$, up to small corrections\footnote{Note that the same parameter~$\Delta^2$ controls both the SNR for the~$n$-point functions and the corrections to the SNR. This is expected in perturbation theory where all nonlinearities are controlled by~$\Delta^2$. Small corrections to Eq.~\eqref{eq:snrest} are captured by the non-Gaussian covariance matrix. The smallness of~$\Delta^2$ justifies why using the simple Gaussian covariance matrix is sufficient on large scales.} of order~$\Delta^2(k_{\rm max})$. One can compute the exact SNR using the standard Gaussian likelihood that we will describe below and check that for higher-order~$n$-point functions the agreement with the estimate above is rather good for small values of nonlinear bias parameters. This is shown in Fig.~\ref{fig:SNR} where the exact calculations for the SNR of P, B and T are compared to Eq.~\eqref{eq:snrest}. Note that the signal-to-noise in each higher order $n$-point function is roughly one order of magnitude smaller than in the previous one, given the value of~$\Delta^2(k_{\rm max})$. This implies that, in the absence of almost perfect degeneracies, the information content of the nonlinear field on large scales is almost saturated by the leading summary statistics. One well-known strong degeneracy for biased tracers in real space is between the amplitude of the linear fluctuations~$A$ and the linear bias~$b_1$. This degeneracy is broken by combining the power spectrum and bispectrum, and hence the total constraining power of real space data is suppressed by $\Delta^2$ compared to the total signal in the power spectrum.

What do our assumptions imply for the FBI analysis? Given the Eulerian description, perturbativity prior and the negligible shot noise, one can compute the posterior for the FBI perturbatively. It is then possible to argue that FBI is equivalent, order by order in perturbation theory, to the analysis based on all $n$-point correlation functions that can be reliably computed up to the given order in the perturbative expansion used for the nonlinear field in the FBI. On the level of equations this equivalence has been explicitly demonstrated in~\cite{Cabass:2023nyo} at leading order in~$\Delta^2$ and more recently in~\cite{Schmidt:2025iwa} at subleading order in~$\Delta^2$ and for the full cubic bias model. We refer the reader to these references for further details. Showing this connection using simulation data is more challenging and beyond the scope of this paper. Here we will ask two concrete simple questions that can still shed some light on the connection between $n$-point functions and FBI. First, we can ask how well the amplitude~$A$ can be measured with P+B combination using a simple model and the usual Gaussian likelihood. Second, using the Fisher matrix, we can ask if addition of the trispectrum to the P+B analysis significantly improves the errors. 

In order to answer these questions, in Sec.~\ref{sec:models} we consistently compute the one-loop power spectrum, the tree-level bispectrum and the tree-level trispectrum for dark matter halos in real space.  We explicitly check that our one-loop power spectrum and tree-level bispectrum model is unbiased, performing the real analysis on the PT Challenge simulation suite~\cite{Nishimichi:2020tvu}\footnote{The power spectrum data for the PT challenge simulations are publicly available \href{https://www2.yukawa.kyoto-u.ac.jp/~takahiro.nishimichi/data/PTchallenge/}{here}, see~\cite{Nishimichi:2020tvu} for more details.}, as shown in Sec.~\ref{sec:simul}. We then perform analyses and forecasts using mock data and measure the amplitude of the linear density field for a simulation box with volume $V=(2\;{\rm Gpc}/h)^3$ and number density of tracers as in~\cite{Nguyen:2024yth} and~\cite{Beyond-2pt:2024mqz}. Our two main findings are as follows. 

\begin{itemize}
\item {\em First, we find that the errors for the amplitude~$A$ in the P+B analysis are: (a) in agreement with the simple mode counting estimate of eq.~\eqref{eq:snrest}, and (b) similar to the errors obtained in~\cite{Nguyen:2024yth} using FBI (and therefore significantly smaller than in the SBI P+B analysis of~\cite{Nguyen:2024yth})}. Our findings are in line with the previous literature on the joint P+B analysis and we demonstrate that our constraints are robust against inclusion of the theoretical error that mimic addition of higher order nonlinearities in the analysis. 
\item {\em Second, we find that, in a generic setup, adding the tree-level trispectrum to the P+B analysis only slightly improves the error on~$A$}. The improvement we find is of the order~$\mathcal{O}(20\%)$ and in agreement with the estimate from eq.~\eqref{eq:snrest}. We use a number of simplifying assumptions about the theoretical model for T and its cross-covariance with P and B, which all maximize the impact of T in the joint analysis.   
\end{itemize}

Our results, presented in Sec.~\ref{sec:results}, indicate that in the simple setup where the only cosmological parameter is the amplitude of the linear field~$A$, the information content in the nonlinear field on large scales is almost saturated with the power spectrum and bispectrum and that P+B+T analysis leads to very similar errors as the FBI of~\cite{Nguyen:2024yth} and~\cite{Beyond-2pt:2024mqz}. These results are in line with general expectations in perturbation theory and explicit derivations presented in~\cite{Cabass:2023nyo,Schmidt:2025iwa}. One remaining puzzle is the origin of the difference between the conventional P+B analysis presented in this paper and SBI method for P+B used in~\cite{Nguyen:2024yth} (see also~\cite{Tucci:2023bag}). We will comment on differences in these two approaches and possible explanations for different results in Sec.~\ref{sec:discussion}. 

In order to perform our analyses, we relied mostly on the well-known results for $n$-point functions of biased tracers. However, we also derive several new results such as the covariance matrices for trispectrum analysis and the detailed model for the noise contributions to the trispectrum. Somewhat orthogonally to the main goals of this paper, we also estimate the information content in the BAO wiggles beyond~$k_{\rm max}=0.12\; h/{\rm Mpc}$, and discuss how much information on~$A$ can come from large displacements at nonlinear scales.

\section{Perturbative description of the halo density field}
\noindent Galaxy density field $\delta_g$ in perturbation theory can be written as a sum of two contributions
\begin{equation}
\delta_g = \delta_g^{\rm model} + \epsilon_g \;. 
\end{equation}
The first term is the model $\delta_g^{\rm model}$ which depends on the initial conditions and cosmological parameters. The second term is the ``noise'' $\epsilon_g$, which captures the stochastic nature of halo and galaxy formation. This contribution can be understood to arise from integrating over small scales to which we do not have access in observations (scales smaller than the typical intergalactic separation). The noise is uncorrelated with the matter density field, but it can be modulated by the long-wavelength modes which leads to several nontrivial contributions to the galaxy bispectrum and trispectrum. In this section we will remind the reader how perturbative descriptions for both $\delta_g^{\rm model}$ and $\epsilon_g$ look like and why the simple Eulerian theory for $\delta_g^{\rm model}$ is sufficient to capture all relevant nonlinearities in real space on large enough scales.  

\subsection{Why is Eulerian description sufficient on large scales?}\label{sec:EPTvsLPT}

\noindent It is well known that Lagrangian perturbation theory (LPT) reproduces very well the nonlinear fields of dark matter or biased tracers, in real and redshift space. Even when using the Zel'dovich approximation only, one can generate realizations with familiar filamentary structures, which correlate well with the simulations. For this reason, LPT is often employed when making perturbative maps or doing detailed comparisons of simulations and perturbation theory. Furthermore, LPT automatically accounts for effects such as broadening of the BAO peak (or other features) and can be easily implemented in practice on a grid. Given all this, it is natural that the field-level analyses use forward modeling based on LPT to evolve a given initial linear density field (provided cosmological, noise and bias parameters). LPT up to cubic order with all relevant bias operators in the initial field was used in~\cite{Nguyen:2024yth}, as well as in the previous works~\cite{Stadler:2023hea,Stadler:2024aff,Stadler:2024fui,Babic:2022dws,Babic:2024wph,Tucci:2023bag}. 

However, most of the field-level inference has been done so far on very large scales. For example, the largest $k_{\rm max}$ cut used in~\cite{Nguyen:2024yth} is $k_{\rm max}=0.12\;h/$Mpc. In this regime, there is a significant simplification in the nonlinear dynamics and Eulerian perturbation theory (EPT) is sufficient to take into account all relevant nonlinearities in the field. While all theoretical calculations can be done in LPT as well, the advantage of EPT is that it is easier to work with, it expresses the nonlinear field in terms of the initial conditions in a simple way and allows to more easily disentangle different physical effects that can affect the analysis (such as displacements and broadening of the BAO peak, versus the growth of structure). Since one of the goals of this paper is to better analytically understand the error bars for the field-level and conventional analyses, we will use EPT. 

Why is EPT sufficiently good on large enough scales? There are two main reasons:
\begin{itemize}
    \item First, the nonlinear fields generated using cubic EPT and LPT up to $k_{\rm max}=0.12\;h/$Mpc and at redshifts $z=0.5$ and $z=1$ are almost indistinguishable, i.e.,~their cross-correlation coefficient is close to 1. This has been verified explicitly at the field level~\cite{Taruya:2018jtk} and can be easily understood analytically. We will review the main arguments below.  
    \item Second, even though the FBI is based on comparing the fields predicted in theory and data, it eventually boils down to measuring some summary statistics due to the averaging over all possible initial conditions, which is equivalent to computing the expectation values~\cite{Cabass:2023nyo,Schmidt:2025iwa}. It is well known that in EPT and LPT the expectation values are in much closer agreement than what the cross-correlation coefficient indicates, making EPT even more adequate in approximating LPT. 
\end{itemize}

Let us review these two arguments in detail. We begin by estimating the cross-correlation coefficient between EPT and LPT. Let us imagine that in the initial (Lagrangian) coordinates $\q$ we have a linear field $\delta_1(\q)$. We can imagine that we evolve this field in two different ways. In EPT, if we use just the linear evolution, the final field is just proportional to $\delta_1(\q)$. On the other hand, in LPT we have to displace particles from their initial coordinates $\q$ to their final Eulerian positions~$\x$, such that $\x=\q+\boldsymbol\psi(\q)$. The leading contribution to the nonlinear displacement field~$\boldsymbol\psi(\q)$ is given by the Zel'dovich approximation $\boldsymbol\psi_1(\q)=-\boldsymbol{\nabla}/\nabla^2 \delta_1(\q)$, such that the leading order (LO) LPT field is given by
\begin{equation}
\delta_{\rm LPT}^{\rm LO}(\k) = \int d^3\q e^{-i\k\cdot\boldsymbol\psi_1(\q)} e^{-i\k\cdot\q} \;.
\end{equation}
Note that here we assume that all quantities are evaluated in linear theory at the final redshift and we suppress explicit time dependence. 


Starting from this equation one can compute the cross-correlation coefficient between the leading order LPT and EPT fields. It is given by
\begin{equation}
\label{eq:cross_linear}
r^{\rm LO}(k) \equiv \frac{\langle \delta_{\rm EPT}^{\rm LO}(\k') \delta_{\rm LPT}^{\rm LO} (\k) \rangle'}{\sqrt{P_{\rm EPT}(k) P_{\rm LPT}(k) }} \approx e^{-\frac{1}{2}k^2 \sigma_v^2} \;,
\end{equation}
where prime on the correlation function indicates that $(2\pi)^3\delta^D(\k+\k')$ has been removed, $P_{\rm EPT}(k)$ and $P_{\rm LPT}(k)$ are leading order power spectra and the velocity dispersion is given by
\begin{equation}
\sigma_v^2 = \frac{1}{6\pi^2} \int_{0}^\infty dq P_{\rm lin} (q) \;.
\end{equation}
Derivation of this expression is given in Appendix~\ref{app:EPTvsLPT}. Let us here provide an intuitive justification for it. The major difference between EPT and LPT is the displacement field. Therefore, on scales larger than the typical variance of the displacement field, given by $\sigma^2_v$, the EPT and LPT fields are the same. On smaller scales they rapidly decorrelate and $r^{\rm LO}(k)$ goes to zero. The value of $\sigma_v^2$ in a $\Lambda$CDM-like cosmology is approximately
\begin{equation}
\sigma_v^2 (z) = 36\left(\frac{D_+(z)}{D_+(0)} \right)^2 \; ({\rm Mpc}/h)^2 \;.
\end{equation}
This implies a significant decorrelation between the leading order EPT and LPT fields, even at $k_{\rm max}=0.1\;h/$Mpc. 

\begin{figure}
    \centering
    \includegraphics[width=0.49\linewidth]{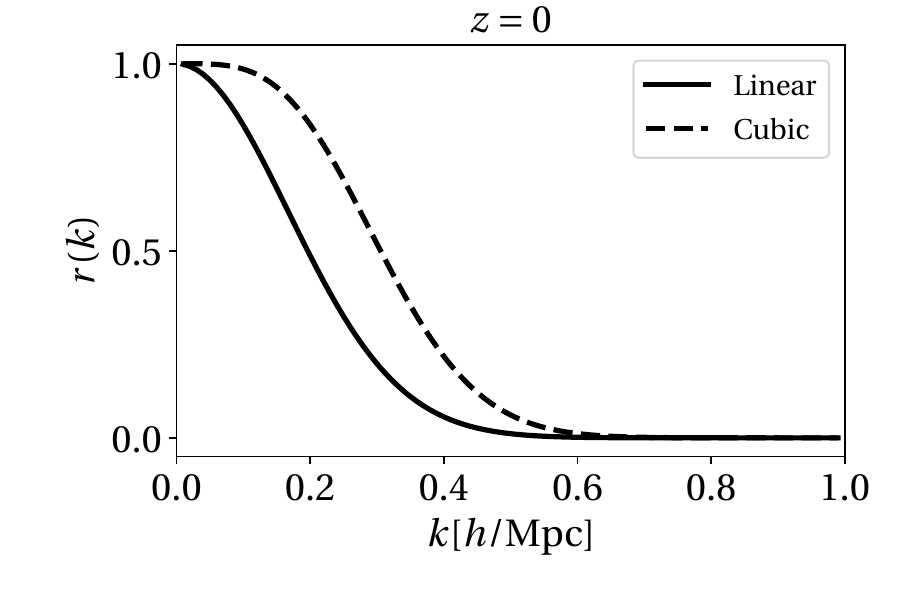}
    \includegraphics[width=0.49\linewidth]{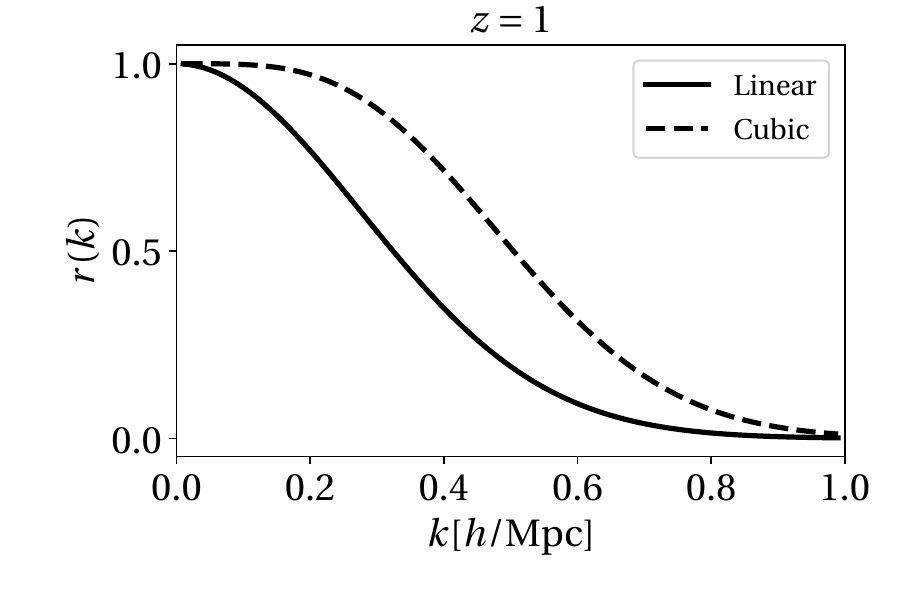}
    \caption{Estimate of the cross correlation coefficient between the EPT and LPT fields in the linear and cubic model at~$z=0$ (left panel) and~$z=1$ (right panel). These curves have the usual ``s'' shape and are in excellent agreement with the findings from the explicit grid computations of EPT and LPT (for example, see Fig.~10 in~\cite{Taruya:2018jtk}). }
    \label{fig:cross_coeff_mat}
\end{figure}

However, in our analysis we will go beyond the leading order approximation. If we use cubic models for both EPT and LPT the agreement between the two improves. In this case a part of the nonlinearities induced by the large displacements is taken into account in EPT. Since the two theories match at one-loop order on large scales, the cross-correlation coefficient at next-to-leading order (NLO) will be better and must scale as 
\begin{equation}
r^{\rm NLO}(k) = 1 + \mathcal{O}(k^4\sigma_v^4) \;,
\end{equation}
in the low-$k$ limit, where $\mathcal{O}(k^4\sigma_v^4)$ is the leading contribution from the unaccounted two-loop terms. While deriving the exact expression is more involved in this case, we can provide a simple estimate of the cross-correlation coefficient at NLO. In a way identical to what happens in the so-called infrared resummation, we can write 
\begin{equation}\label{eq:cross_cubic}
r^{\rm NLO}(k) \approx \left(1 + \frac{1}{2}k^2 \sigma_v^2 \right) e^{-\frac{1}{2}k^2 \sigma_v^2} \;,   
\end{equation}
which is guaranteed to produce the correct low-$k$ limit for the cubic EPT and LPT models, while keeping the same exponential suppression on small scales. In Fig.~\ref{fig:cross_coeff_mat} we plot the cross-correlation coefficient for linear and cubic model, corresponding to~Eq.~\eqref{eq:cross_linear} and Eq.~\eqref{eq:cross_cubic} respectively. Our estimates agree very well with measurements of the cross-correlation coefficient from explicit EPT and LPT maps built at different order in perturbation theory (for example, see Fig.~10 in~\cite{Taruya:2018jtk}). Going to cubic order in EPT and having the correct one-loop power spectrum, makes the fields correlate much better.

\begin{figure}
    \centering
    \includegraphics[width=0.49\linewidth]{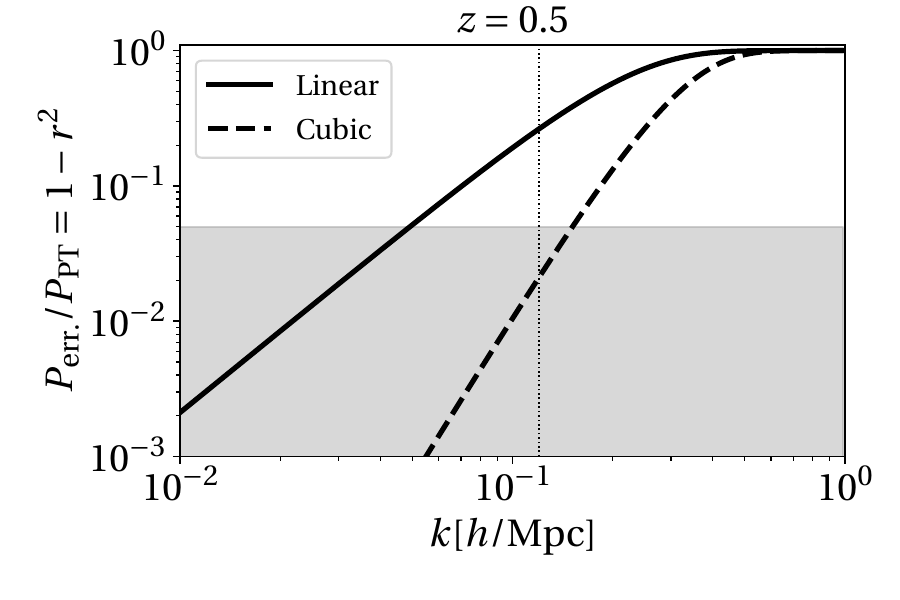}
    \caption{Power spectrum of the difference between LPT and EPT fields at~$z=0.5$ in linear and cubic models, normalized by~$P_{
    \rm PT}$, see Eq.~\eqref{eq:Perr_DM}. The shaded region corresponds to a relative error smaller than~$5\%$. Note that~$P_{\rm err.}/P_{\rm PT}\approx 2 \Delta A/ A$. Thin vertical line marks the maximum wavenumber used in the analysis,~$k_{\rm max}=0.12\;h/{\rm Mpc}$. At higher redshifts the agreement between EPT and LPT is even better.}
    \label{fig:cross_coeff_mat_err}
\end{figure}

The cross-correlation coefficient is related to the error that one makes approximating LPT with EPT. If we define the ``error’’ field as the difference~$\delta_{\rm err} \equiv \delta_{\rm LPT}-\delta_{\rm EPT}$, then the power spectrum of this difference can be approximated as follows
\begin{equation}
\label{eq:Perr_DM}
P_{\rm err.} = \langle |\delta_{\rm LPT} - \delta_{\rm EPT} |^2 \rangle’ = P_{\rm LPT} + P_{\rm EPT} + 2\sqrt{P_{\rm LPT}P_{\rm EPT}} r \approx P_{\rm PT} (1-r^2) \;,
\end{equation}
where in order to make the last approximation we used that $P_{\rm LPT}\approx P_{\rm EPT} = P_{\rm PT}$ and that~$r\approx 1$ on large scales, such that~$2(1-r)\approx (1+r)(1-r)=1-r^2$. In other words, the quantity~$1-r^2$ measures the relative size of the error power spectrum compared to the~$P_{\rm PT}$. In Fig.~\ref{fig:cross_coeff_mat_err} we plot~$1-r^2$ for redshift~$z=0.5$. We can see that the relative error of approximating LPT with EPT is at the percent level for the scale cuts used in the analyses,~$k_{ \rm max}=0.1 \; h/{\rm Mpc}$ and~$k_{ \rm max}=0.12 \; h/{\rm Mpc}$. This is much smaller than the~$\mathcal O(5\%)$ relative error on the amplitude of the linear field that can be obtained from a~$(2\; {\rm Gpc}/h)^3$ volume box of biased tracers in real space~\cite{Nguyen:2024yth}. This implies that on large scales and within given statistical errors, Eulerian and Lagrangian perturbative forward models are practically indistinguishable.

\begin{figure}
    \centering
    \includegraphics[width=0.49\linewidth]{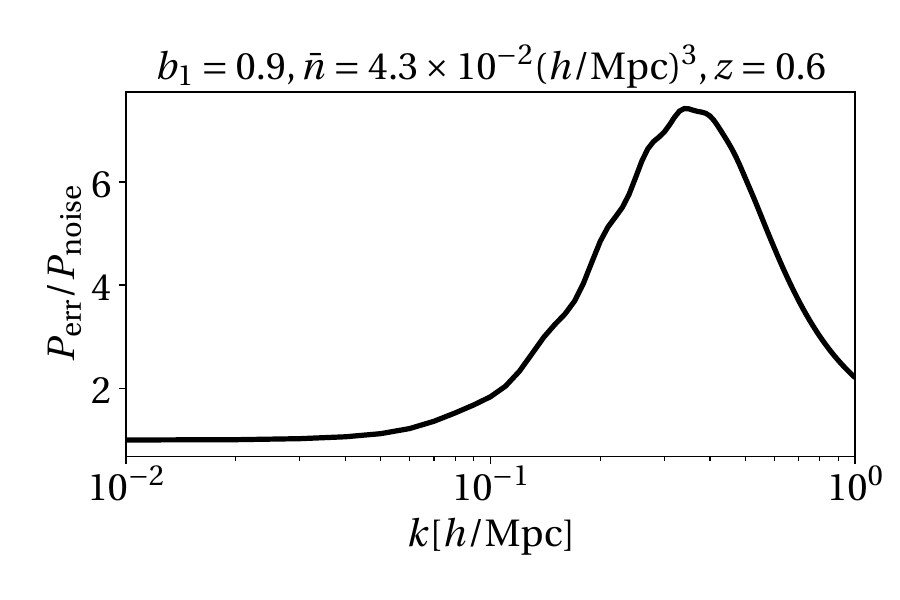}
    \includegraphics[width=0.49\linewidth]{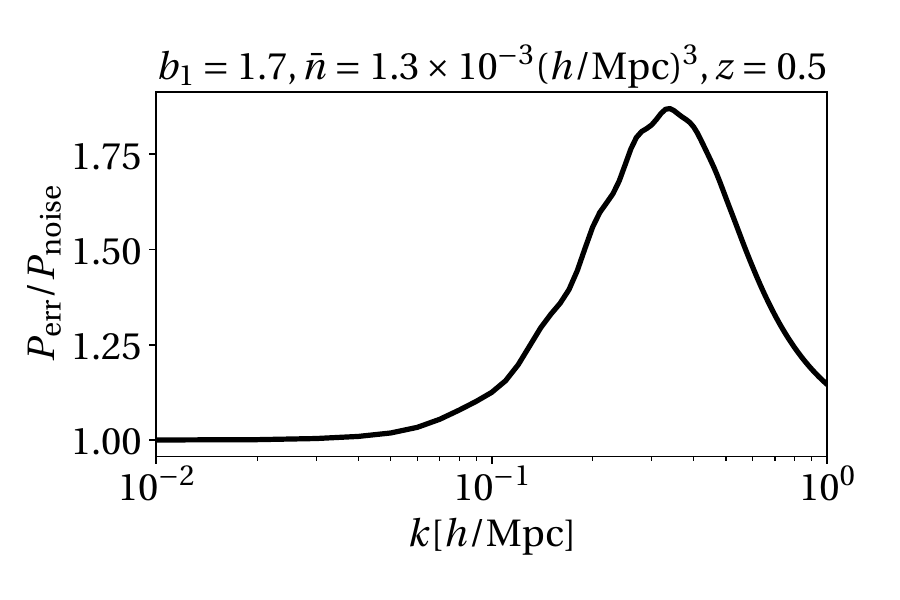}
    \caption{$P_{\mathrm{err}}/P_{\mathrm{noise}}$ as function of scale for two biased tracers similar the ones considered in~\cite{Schmittfull:2018yuk} (\textit{Left)} and in~\cite{Nguyen:2024yth} (\textit{Right}). }
    \label{fig:perr}
\end{figure}

One may object that the arguments we presented so far apply only to the dark matter field. However, the situation for the biased tracers is in some sense even better, thanks to the additional source of the error due to the discrete sampling of the halo density field. Assuming constant shot noise power spectrum on large scales with the amplitude~$P_{\rm noise}$, we can write the ratio of the total error\footnote{For galaxies we assume~$\delta_{g,\rm LPT}-\delta_{g,\rm EPT} = \delta_{\rm err.}+\epsilon_g$, where $\langle \epsilon_g \epsilon_g \rangle' = P_{\rm noise}$.} power spectrum to the noise as follows
\begin{equation}
\frac{P_{\rm err.}}{P_{\rm noise}} \approx 1 + \left(1 - r^2 \right) \frac{b_1^2 P_{\rm lin}(k)}{P_{\rm noise}} \;.
\end{equation}
As expected, for biased tracers with large enough shot noise, all modeling errors are negligible compared to the shot noise $P_{\rm noise}$ on large scales, when~$r$ goes to zero. In the left panel of Fig.~\ref{fig:perr} we plot the ratio from the previous formula for a very dense tracers with $\bar{n}\sim4.3\times10^{-2}(h/\mathrm{Mpc})^3$. This agrees very well with measurements from simulations (for example, see the dark orange line on the left panel of Fig.~(20) in~\cite{Schmittfull:2018yuk}). In the right panel of Fig.~\ref{fig:perr} we show the same ratio but for the tracer considered in~\cite{Nguyen:2024yth}. While the modeling error is roughly~$10\%$ larger than the sampling noise at~$k_{\rm max}=0.12\;h/{\rm Mpc}$, the sampling noise itself is just a small fraction of the total power spectrum on these scales. Therefore, the cross-correlation coefficient between EPT and LPT can be very easily understood both for dark matter field and biased tracers. In both cases the two perturbative approaches agree at the percent level on relevant scales.

Let us also briefly discuss the second reason for why EPT is a good approximation for LPT on large scales. So far we estimated the cross correlation coefficient and the error power spectrum and showed that on scales relevant for our analysis the EPT and LPT fields match at the percent level. However, in the FBI one performs the marginalization over all initial conditions, resulting in effective measurement of some summary statistics from the maps. In the perturbative regime these statistics are just the leading $n$-point functions~\cite{Cabass:2023nyo}. Regardless of the details, we want to argue that the estimate of the mismatch based on the cross-correlation coefficient is the most pessimistic one. It is dominated by the velocity dispersion induced by the large bulk flows. However, it is well known that due to the equivalence principle the effects of these bulk flows exactly cancel in all summary statistics (with the exception of features such as BAO, which is irrelevant on large scales)~\cite{Peloso:2013zw,Kehagias:2013yd,Creminelli:2013poa,Creminelli:2013mca,Senatore:2014via,Mirbabayi:2014gda,Baldauf:2015xfa}. Therefore, the agreement between EPT and LPT predictions can be only better in practice than the~$1-r^2$ estimate indicates. For example, even though the Zel'dovich and linear Eulerian fields strongly decorrelate for~$k\gtrsim 1/\sigma_v$, the two power spectra do not differ dramatically. The same is true for any higher order correlation function or other statistical quantities. 

In conclusion, for the range of scales and redshifts used in~\cite{Nguyen:2024yth} and our analysis in this paper, the standard Eulerian perturbation theory (without the IR resummation) and Lagrangian perturbation theory describe the halo density field equally well. Since there is no distinction between the two nonlinear fields on the level of summary statistics, all conclusions based on EPT must be valid for LPT as well.

\subsection{Halo density field in Eulerian perturbation theory}

\noindent Given the arguments that justify using the EPT, we now turn to the model for the galaxy density field. 
In EPT the nonlinear density field of biased tracers in real space on large scales has a simple form in terms of local\footnote{Since dark matter halo formation happens over a long time with the characteristic scale~$t_{\rm gal.~form.}\sim H^{-1}$, the overdensity of tracers at a given time depend on the entire history of the long-wavelength dark matter density field evolution, making the most general bias expansion non-local in time~\cite{Tegmark:1998wm,Senatore:2014eva,Mirbabayi:2014zca,Chan:2012jj, Donath:2023sav}. However, it has been shown that up to cubic order these non-local contribution can be rewritten such that the expansion \eqref{eq:bias_expansion_real} looks local.} operators~\cite{Desjacques:2016bnm}:
\begin{equation}
    \delta_g^{\rm model}(\x) = b_1 \delta(\x) + \frac{b_2}{2} (\delta^2(\x)-\langle \delta^2 \rangle) + b_{\mathcal{G}_2}\mathcal{G}_2(\x) + \frac{b_3}{6}\delta^3(\x) + b_{\mathcal{G}_3}\mathcal{G}_3(\x) + b_{\mathcal{G}_2 \delta} \mathcal{G}_2(\x)\delta(\x) + b_{\Gamma_3} \Gamma_3(\x) - c_s^2 \frac{\nabla^2}{k_{\rm NL}^2} \delta(\x) \ \;,
    \label{eq:bias_expansion_real}
\end{equation}
where $\delta$ is the nonlinear matter field (to be computed in Eulerian perturbation theory up to cubic order) and bias operators $\mathcal{G}_2$, $\mathcal{G}_3$ and $\Gamma_3$ are built from tidal fields of gravitational or velocity potential. These terms are sufficient to describe the one-loop power spectrum, tree-level bispectrum and tree-level trispectrum and they match the cubic Lagrangian perturbation theory. Note that the last contribution in Eq.~\eqref{eq:bias_expansion_real} is a higher derivative term, which has a different expansion parameter, given by $k/k_{\rm NL}$ and not by the variance of $\delta$. For a $\Lambda$CDM-like power spectrum this term is comparable to other cubic contributions and has to be kept. It is important to remember that there are two physically distinct effects that are described by the higher derivative operator. One is the response of dark matter halos to $\nabla^2\delta$ and the other one is the dark matter one-loop counterterm. Since at this order in perturbation theory these two effects are completely degenerate, we will keep only one free parameter in our analysis.\footnote{{In principle, there are two different scales associated to the two different effects. The dark matter counterterm has the nonlinear scale $k_{\rm NL}$, while the dark matter halos have another scale $k_*$ related to the Lagrangian size of a typical halo in the sample. For massive halos these two can be very different, but we will have in mind dense samples as in~\cite{Nguyen:2024yth} where the two scales are comparable. Anyway, since we assume that the higher derivative bias and the counterterm are both unknown, this simplification does not impact our analysis.}}

In summary, for the purposes of this work, we will use eight unknown parameters that depend on the halo or galaxy properties and have to be measured from the data. These are 
\begin{equation}
\theta_{\rm bias} = \{b_1, b_2, b_{\mathcal G_2}, b_3, b_{\mathcal G_3}, b_{\mathcal G_2 \delta}, b_{\Gamma_3}, c_s^2 \} \;.
\end{equation}
Biases and counterterms are (strongly) degenerate with some cosmological parameters and measuring multiple correlation functions is crucial to break these degeneracies. In our analysis, we will focus only on one cosmological parameter which measures the amplitude of the linear density fluctuations $A$. We define the following parameter
\begin{equation}
\alpha \equiv \frac{A}{A^{\rm fid}} = \frac{\sigma_8}{\sigma_8^{\rm fid}} \;, 
\end{equation}
where the second equality holds if all other cosmological parameters are fixed as in~\cite{Nguyen:2024yth}.

We will do all calculations and the analysis in Fourier space. Given the real space model and parameters, we can write 
\begin{equation}
\delta_g^{\rm model} (\k,z) = \sum_n \alpha^n D^n(z) \int_{\q_1} \cdots \int_{\q_n} (2\pi)^3 \delta^D(\k-\q_{1\ldots n}) K_n(\q_1,\ldots,\q_n;\theta) \delta_1(\q_1) \cdots \delta_1(\q_1) - \alpha c_s^2 \frac{k^2}{k_{\rm NL}^2} D(z)\delta_1(\k) \;,
\end{equation}
where $\int_{\q} \equiv \int d^3\q/(2\pi)^3$, $\q_{1\ldots n}\equiv \q_1+\cdots \q_n$, $D(z)\equiv D_+(z)/D_+(0)$ is the normalized growth factor at redshift $z$ such that $D(0)=1$, $\delta_1(\k)$ is the linear density field and $K_n$ are perturbation theory kernels that contain all information about the nonlinear evolution. These kernels are well-known in the literature (see for example ~\cite{Assassi:2014fva,Senatore:2014eva,Angulo:2015eqa, DAmico:2021rdb}) and here we just write them down for completeness. They are given as
\begin{equation}
K_1(\q;\theta) = b_1 \;,
\end{equation}
\begin{equation}
K_2(\q_1,\q_2;\theta) = b_1 F_2(\q_1,\q_2) + \frac{b_2}{2} + b_{\mathcal G_2} \left( \frac{(\q_1\cdot\q_2)^2}{q_1^2q_2^2} -1 \right) \;,
\end{equation}
\begin{align}
K_3(\q_1, & \q_2,\q_3;\theta) = b_1 F_3(\q_1,\q_2,\q_3) \nonumber \\
& + \frac{b_2}{3}(F_2(\q_1,\q_2) + F_2(\q_1,\q_3) + F_2(\q_2,\q_3) ) + 2b_{\mathcal G_2} F_{\mathcal G_2}(\q_1,\q_2,\q_3) \nonumber \\
& + \frac{b_3}{6} + b_{\mathcal G_3} F_{\mathcal G_3} (\q_1,\q_2,\q_3) + b_{\Gamma_3} F_{\Gamma_3} (\q_1,\q_2,\q_3) + \frac{b_{\mathcal G_2\delta}}{3} \left( \frac{(\q_1\cdot\q_2)^2}{q_1^2q_2^2} + \frac{(\q_1\cdot\q_3)^2}{q_1^2q_3^2} + \frac{(\q_2\cdot\q_3)^2}{q_2^2q_3^2} - 3 \right) \;.
\end{align}
In these expressions $F_2$ and $F_3$ are the standard perturbation theory kernels, see~\cite{Bernardeau:2001qr}, and additional terms are defined by
\begin{equation}
F_{\mathcal G_2}(\q_1,\q_2,\q_3) \equiv  \frac1{3} F_2(\q_1,\q_2) \left( \frac{((\q_1+\q_2)\cdot\q_3)^2}{|\q_1+\q_2|^2q_3^2} -1 \right) + 2\;{\rm perm} \; .
\end{equation}
\begin{equation}
F_{\mathcal G_3}(\q_1,\q_2,\q_3) \equiv - \frac1{2} \left( 1-  \frac{(\q_1\cdot\q_2)^2}{q_1^2q_2^2} - \frac{(\q_1\cdot\q_3)^2}{q_1^2q_3^2} - \frac{(\q_2\cdot\q_3)^2}{q_2^2q_3^2} + 2 \frac{(\q_1\cdot\q_2)(\q_1\cdot\q_3)(\q_2\cdot\q_3)}{q_1^2q_2^2q_3^2} \right) \; ,
\end{equation}
\begin{equation}
F_{\Gamma_3}(\q_1,\q_2,\q_3) \equiv - \frac4{21} \left( \frac{(\q_1\cdot\q_2)^2}{q_1^2q_2^2} -1 \right) \left( \frac{((\q_1+\q_2)\cdot\q_3)^2}{|\q_1+\q_2|^2q_3^2} -1 \right) + 2\;{\rm perm} \; .
\end{equation}

The kernels $K_n$ can be used to compute the one-loop power spectrum, tree-level bispectrum and tree-level trispectrum. We will do that in the next section. Before that, we discuss the second important ingredient which is the stochastic noise.

\subsection{Stochastic noise}
\noindent As we already explained, the galaxy density field in perturbation theory is not fully deterministic, but it also contains stochastic noise which comes from our ignorance of the small-scale modes that determine galaxy formation and that are not a part of observations (or perturbation theory). This noise does not correlate with long-wavelength density fluctuations, but it can be modulated by the long modes. Therefore, the noise field can be written in a similar manner as the bias expansion
\begin{equation}
\label{eq:noisemodel}
\epsilon(\x)= \epsilon_0(\x) + \epsilon_1(\x) \delta(\x) + \frac 12 \epsilon_2(\x)(\delta^2(\x)-\langle \delta^2 \rangle) + \epsilon_{\mathcal G_2}(\x) \mathcal G_2 (\x) + \ldots \;.
\end{equation}
Note that $\epsilon_i(\x)$ are non-Gaussian fields that can have significant three-point functions.
It is clear from this expression that $\langle \delta \epsilon \rangle =0$ as expected. The auto power spectrum of the noise does not inherit the ``shape'' of the cross spectra from $\delta$,  $\delta^2$ or $\mathcal G_2$ due to the fact that the stochastic fields correlate only in the same spatial point
\begin{equation}
\langle \epsilon_i(\x) \epsilon_j(\x') \rangle \sim c_{ij} \delta^D(\x-\x') \;.
\end{equation}
On large scales the noise power spectrum is approximately constant, with the amplitude approximately equal to the Poisson prediction
\begin{equation}
P_{\epsilon\epsilon}(k) = \frac{c_1}{\bar n_g} \;.
\end{equation}
where $c_1$ is a free dimensionless parameter of order one. When computing higher order $n$-point functions there are several additional terms that may appear and we discuss them in the next section.

\section{Galaxy power spectrum, bispectrum and trispectrum}\label{sec:models}

\noindent Equipped with the perturbative forward model and the noise model, we are now ready to compute all relevant correlation functions. In this section we write down the one-loop power spectrum, a simple model for the BAO damping that we will use in the BAO analysis, tree-level bispectrum and tree-level trispectrum.

\subsection{Power spectrum}\label{sec:pk_model}

\noindent The power spectrum of biased tracers in real space up to one-loop corrections is given by 
\begin{align}
    P_{g}(k,z) = & \; b_1^2\alpha^2 P_{\rm lin}(k,z) +  b_1^2 \alpha^4 P_{13}(k,z) + b_1^2\alpha^4 P_{22}(k,z) -b_1^2c_s^2 \alpha^2 k^2P_{\rm lin} (k,z) \nonumber\\
    & + b_1 b_2 \alpha^4 \mathcal{I}_{\delta^2}(k,z) + \frac{b_2^2}{4} \alpha^4 \mathcal{I}_{\delta^2 \delta^2}(k,z) + b^2_{\mathcal{G}_2} \alpha^4 \mathcal{I}_{\mathcal{G}_2\mathcal{G}_2}(k,z) + \frac{b_2 b_{\mathcal{G}_2}}{2} \alpha^4 \mathcal{I}_{\delta^2 \mathcal{G}_2}(k,z)\nonumber\\
    & + \left(2 b_1 b_{\mathcal{G}_2} + \frac{4}{5} b_1 b_{\Gamma_3}\right) \alpha^4 \mathcal{F}_{\mathcal{G}_2}(k,z) +2 b_1 b_{\mathcal{G}_2} \alpha^4 \mathcal{I}_{\mathcal{G}_2}(k,z) +\frac{c_1}{\bar n_g} \;,
\end{align}
where the explicit form of different contributions can be found in~\cite{Assassi:2014fva,Chudaykin:2020aoj}.
All nuisance parameters in this expression are functions of time, but for the fixed redshift they can be treated as constants. This form of the power spectrum is particularly convenient since all nuisance and cosmological parameters enter the model just as simple multiplicative factors. At the level of the linear theory, $b_1$ and $\alpha$ are perfectly degenerate and only the combination $b_1\alpha$ can be measured with very high precision. This degeneracy can be in principle broken only by the loop contributions, but those have much smaller signal-to-noise, suppressed by the variance of the density field at the maximum wavenumber used in the analysis. In addition, since there are several new nuisance parameters at the loop level, the amplitude~$\alpha$ is expected to be poorly constrained from the P alone. We will verify this explicitly below. 

Let us stress that the correct model for the galaxy power spectrum contains the infrared-resummation, which correctly accounts for the broadening of the BAO peak. However, we will not use it in our baseline analysis for several reasons. First, in line with the conclusions of the previous section, we expect that the infrared-resummation cannot significantly affect the power spectrum on scales and redshifts that we use. Second, we will test this simple model against large-volume PT challenge simulations suit and explicitly show the validity of our approximation. Finally, even if this approximation was to miss a small, percent level effects, that would be irrelevant for the conclusions regarding the expected errors form a joint P+B and P+B+T analyses. 

Clearly, we can neglect the infrared resummation only on very large scales. On small scales, there is an additional information about the amplitude of the linear density field in the broadening of the BAO wiggles. While this is not included in the analysis of~\cite{Nguyen:2024yth} and it is not the main goal of this paper, we explore how much information on~$\alpha$ can be extracted from small scales using BAO. We next give a simple model of the IR-resummation at small scales appropriate for our forecast.

\subsection{BAO wiggles}\label{sec:BAO_model}

\noindent One nontrivial way in which the amplitude $\alpha$ affects the BAO is through the broadening of the BAO peak. This nonlinear effect is very well understood and can be described to very high precision in perturbation theory~\cite{Senatore:2014via,Mirbabayi:2014gda,Baldauf:2015xfa,Senatore:2017pbn}. First, one can decompose the linear power spectrum into the smooth and wiggly parts~\cite{Baumann:2017gkg,Hamann:2010pw, Vlah:2015zda,Chudaykin:2020aoj};
\begin{equation}
P_{\text{lin}}(k,z) = P_{\rm lin}^{\rm nw}(k,z) + P_{\rm lin}^{\rm w}(k,z)\;. 
\end{equation}
The bulk flows induced by the long wavelength fields affect the features such as BAO, leading to the damping of the oscillations. The full expression for the infrared-resummed galaxy power spectrum is schematically given by
\begin{equation}
\label{eq:FullIRres}
P_g^{\rm IR}(k,z) = P_g^{\rm nw}(k,z) + e^{-\alpha^2\Sigma_\Lambda^2(z)k^2}(1+\alpha^2\Sigma_\Lambda^2(z)k^2) \alpha^2 b_1^2 P_{\rm lin}^{\rm w} (k,z) + e^{-\alpha^2\Sigma_\Lambda^2(z)k^2} P_{g,{\rm 1-loop}}^{\rm w} (k,z) \;, 
\end{equation}
where~$P_g^{\rm nw}$ is the standard EPT power spectrum up to one loop computed using~$P_{\rm lin}^{\rm nw}$ and~$P_{g,{\rm 1-loop}}^{\rm w}$ is the wiggly part of the EPT one-loop power spectrum. The BAO dumping factor is given by~\cite{Baldauf:2015xfa}
\begin{equation}
\label{eq:sigma}
\Sigma_\Lambda^2(z) \equiv \frac{1}{6\pi^2} \int_0^{\Lambda} dq \, P_{\rm lin}(q,z) \left[ 1 - j_0(q\ell_{\rm BAO}) + 2j_2 ( q \ell_{\rm BAO} ) \right]\;, 
\end{equation}
where the BAO scale is $\ell_{\rm BAO} \approx 110 \, h/\text{Mpc}$, $j_n(x)$ are spherical Bessel functions of order \(n\), and \(\Lambda\) is the scale up to which the long-wavelength displacements are taken into account. The typical value used in practice is $\Lambda = 0.2 \, h/\text{Mpc}$, but the result is almost insensitive to this choice~\cite{Baldauf:2015xfa,Blas:2016sfa,Chudaykin:2020aoj}. Given the numerical values of $\Sigma_\Lambda^2(z)$ for relevant redshifts, the damping is important only for high $k$. Note that the formula for the wiggly part of the galaxy power spectrum is {\em non-perturbative} and almost perfectly describes the BAO. Therefore, unlike the broadband, it can be safely used at all scales. 

It is clear from Eq.~\eqref{eq:FullIRres} how the parameter~$\alpha$ can be measured from the BAO wiggles. Apart from the small loop corrections controlled by the perturbation theory parameter~$\Delta^2(k)$,~$\alpha$ multiplies the large parameter~$\Sigma_\Lambda^2$ which controls the broadening of the BAO peak. In other words, the amplitude of the linear density field can be simply estimated from how much the nonlinear evolution damps the BAO wiggles. This is easy to intuitively understand in real space. On large scales, one can approximate the two-point correlation function using only the linear theory. From the smooth part one can measure the combination~$b_1\alpha$. Then, focusing on the BAO peak only, the broadening is controlled by~$\alpha^2\Sigma_\Lambda^2$ which has only one free parameter. This is the essence of the statement that large displacements can be used to measure~$\alpha$ without strong sensitivity on nonlinear galaxy bias if a feature is present in the initial conditions. Crucially, this is true in Fourier space only for high-$k$ part of the BAO wiggles, beyond~$k_{\rm max}\approx0.2\;h/{\rm Mpc}$.

It is possible to test how much information on~$\alpha$ comes from the BAO wiggles using the formulas above. While one can use Eq.~\eqref{eq:FullIRres} directly, just to keep the discussion as simple as possible, we will use in our forecast the following simplified model for the IR-resummed wiggly part of the galaxy power spectrum 
\begin{equation}
    P_g^{\rm w}(k,z)=e^{-\alpha^2\Sigma_{\Lambda}^2(z)k^2} \alpha^2 b_1^2 P_{\rm lin}^{\rm w}(k,z) + c_{\rm w}^2 \alpha^2 b_1^2 k^2 P_{\rm lin}^{\rm w}(k,z) \; . 
\end{equation}
This equation phenomenologically captures all features of the exact result and in a Fisher forecast it is simpler to implement and interpret. We introduce a single parameter~$c_{\rm w}^2$ that encompasses all nonlinear bias and possible other nonlinear corrections and it is of order~$\mathcal O(1)$. The wiggly part of the galaxy power spectrum at high~$k$ can then be analyzed separately from the broadband (see for example~\cite{Philcox:2020vvt,Zhang:2021yna,Chen:2021wdi,Gil-Marin:2022hnv}).

We will consider two different setups. 
\begin{itemize}
    \item Optimistic: In this setup we set~$c_{\rm w}=0$ and~$\Lambda\to \infty$. This is the best case scenario, since we neglect all nonlinear corrections with bias parameters and we replace~$\Sigma_\Lambda^2$ with~$\Sigma_\infty^2$, to approximately take into account the total broadening from the displacement {\em and} loop corrections. This makes the effect as large as possible and the constraints as good as possible.  
    \item Realistic: In this setup we set~$\Lambda=0.2\;h/{\rm Mpc}$ and leave~$c^2_{\rm w}$ as a free parameter with a Gaussian prior of width equal to 1. In this way the information on~$\alpha$ comes only from the part of the broadening induced with large displacements and it is to some extent degenerate with other nonlinearities. 
\end{itemize}

\subsection{Bispectrum and trispectrum}
\noindent Next, we describe the higher order~$n$-point functions that can be consistently computed using the cubic model. The tree-level bispectrum in real space is given by~\cite{Bernardeau:2001qr}:
\begin{equation}
B_g(k_1,k_2,k_3,z)=2b_1^2\alpha^4 K_2(\k_1,\k_2)P_{\rm lin}(k_1,z) P_{\rm lin}(k_2,z) + 2\,\text{perms.} +\frac{c_2}{\Bar{n}_g} b_1 \alpha^2 P_{\rm lin}(k_1,z) + 2\,\text{perms.} +\frac{c_3}{\Bar{n}_g^2}\;.
\end{equation}
Note that there are two additional coefficients in the noise model compared to the power spectrum. These three free parameters~$c_1$,~$c_2$ and~$c_3$ exactly match the noise parameters used in the analysis of~\cite{Nguyen:2024yth}.

The only additional~$n$-point correlation function that can be fully computed using cubic order Eulerian kernels is the connected tree-level trispectrum, which is given by
\begin{equation}
\begin{aligned}
T_g^c(\k_1, \k_2, \k_3, \k_4,z) = & \; 4b_1^2 \alpha^6 P_{\rm lin}(k_1,z) P_{\rm lin}(k_2,z) \big( K_2(\k_1, -\k_{13}) K_{2}(\k_2, \k_{13}) P_{\rm lin} (k_{13},z) \\
&  + K_{2}(\k_1, -\k_{14}) K_{2}(\k_2, \k_{14}) P_{\rm lin}(k_{14},z) \big) + 5\, \text{perm.} \\
& + 6 \, b_1^3 \alpha^6 K_3(\k_1, \k_2, \k_3) P_{\rm lin}(k_1,z)P_{\rm lin}(k_2,z) P_{\rm lin}(k_3,z)+ 3 \, \text{perm.}\\
&+ \frac{c_2}{\Bar{n}_g} B_{mgg}(\k_{12},\k_3,\k_4,z) + 5\, \text{perm.}\; , 
\end{aligned}
\end{equation}
where we have defined 
\begin{equation}
\begin{aligned}
    B_{mgg}(\k_{12}, \k_3, \k_4,z) &\equiv \langle\delta_m(\k_{12},z) \delta_g(\k_3,z) \delta_g(\k_4,z) \rangle' = \\
    & = 2 b_1^2 \alpha^4 F_2(\k_3, \k_4) P_{\rm lin}(k_3,z) P_{\rm lin}(k_4,z) \\
    & \quad + 2b_1 \alpha^4  K_2(\k_{12}, \k_4) P_{\rm lin}(k_4,z) P_{\rm lin}(k_{12},z) + 2b_1 \alpha^4 K_2(\k_{12}, \k_3) P_{\rm lin}(k_3,z) P_{\rm lin}(k_{12},z)\;.
\end{aligned}
\end{equation}
Note that we keep only the largest noise contribution which at leading order has the same parameter~$c_2$ as the bispectrum. We do this for two reasons: (a) in order to be able to have a fair comparison to the field-level analysis of~\cite{Nguyen:2024yth} which does not use the most general noise model, and (b) in order to get the most optimistic results for additional constraining power of the trispectrum. Other (subdominant) noise contributions to the trispectrum can be derived from the noise model in Eq.~\eqref{eq:noisemodel} and we do this in Appendix~\ref{app:noise}. Even though their impact is not expected to be large (given reasonable priors on the amplitude of the noise), they would lead to slightly larger error bars in the trispectrum analysis.

\section{Likelihoods}\label{sec:likelihood}

\noindent In this section we present the the likelihoods for the power spectrum, bispectrum, trispectrum and BAO wiggles that we use in our analyses. Given a very simple form of the likelihood in the EPT we also provide a simple mode counting argument for the expected joint power spectrum and bispectrum analysis. 

\subsection{Power spectrum}
\noindent The power spectrum likelihood on large scales can be approximated by the well-known equation
\begin{equation}
    \ln\mathcal{L}_P=-\frac{1}{2}\sum_{k} \frac{(P_g(k,\theta)-\hat P_g(k))^2}{\text{Cov}_P(k)}\;.
\end{equation}
Note that we assume the Gaussian form of the likelihood which is appropriate on large scales. The sum runs over all~$k$-bins up to a given~$k_{\rm max}$. The set of variables~$\theta$ contains all nuisance and cosmological parameters in the analysis, while the measured power spectrum in a given~$k$-bin is~$\hat P_g(k)$. Note that for simplicity we neglect the explicit redshift dependence. 

From the way we write the power spectrum likelihood it is clear that we assume a simple diagonal covariance matrix, whose elements are given by
\begin{equation}
\text{Cov}_P(k)=\frac{4\pi^2}{V \Delta k k^2}P_g^2 (k,\theta_{\rm fid})\;,
\end{equation}
where~$V$ is the volume of the survey,~$\Delta k$ is the width of the~$k$ bins and~$\theta_{\rm fid}$ is the fiducial value of cosmological and nuisance parameters. On large scales used in our analysis, it is sufficient to use the linear theory in the computation of the covariance. All nonlinear corrections and off-diagonal elements in the covariance matrix are small on large scales and suppressed by the variance of the density field~$\Delta^2(k_{\rm max})$. Furthermore, since we are interested in posterior on cosmological parameters after marginalization over all nuisance parameters, the final variance of cosmological parameters is not impacted only by the data covariance, but also the effective covariance due to the marginalization. This was explicitly demonstrated in some simple examples in~\cite{Chudaykin:2020hbf}. This additional uncertainty is usually much bigger than any mistake made in the data covariance. For this reason, one recovers identical error bars on cosmological parameters on large scales in analyses based on perturbation theory, regardless of the choice of the data covariance. The explicit demonstration of this fact can be found in~\cite{Wadekar:2020hax}. 

\subsection{Bispectrum}
\noindent Similarly to the power spectrum, we assume a simple Gaussian likelihood with the diagonal covariance matrix for the bispectrum as well
\begin{equation}
    \ln\mathcal{L}_B=-\frac{1}{2}\sum_{\mathcal T} \frac{(B_g(k_1,k_2,k_3,\theta)-\hat B_g(k_1,k_2,k_3))^2}{\text{Cov}_B(k_1,k_2,k_3)}\;,
\end{equation}
where the sum now runs over all triangles in Fourier space. In the Gaussian approximation only the same shape triangles correlate and the diagonal elements of the covariance matrix are given by
\begin{equation}
    \text{Cov}_{B}(k_1,k_2,k_3)=\frac{(2\pi)^6}{V V_{123}}s_{123}P_g(k_1,\theta_{\rm fid}) P_g(k_2,\theta_{\rm fid}) P_g(k_3,\theta_{\rm fid})\;,
\end{equation}
where $s_{123}$ is a symmetry factor and it is equal to $6,2,1$ for equilateral, isosceles and scalene triangles respectively and $V_{123}$ is the volume of a given triangle bin which in the thin shell approximation reads
\begin{equation}
V_{123}=8\pi^2k_1k_2k_3\Delta k^3\;.
\end{equation}
Note that we assume ordered momenta, $k_3\leq k_2\leq k_1$, in order to avoid the double counting of equivalent triangles. 

For reasons similar to the power spectrum case, one can justify the simple approximation for the bispectrum likelihood in perturbation theory. Corrections to the Gaussian approximation are expected to be of order~$\Delta^2(k_{\rm max})$ and cannot significantly change the results in the full analysis. We will comment more on this and the cross-covariance with the power spectrum below. 

\subsection{Trispectrum}

\noindent While the form of the power spectrum and bispectrum likelihoods in the Gaussian approximation is well-known, the trispectrum is much less used and explored in the comparisons to simulations and data (for some recent progress, see~\cite{Gualdi:2020eag,Gualdi:2022kwz}). First of all, the connected trispectrum is a function of six scalar variables. It can be parametrized in different ways, but we will use four magnitudes of the wavenumbers~$k_1,\ldots,k_4$ and two magnitudes of the diagonals,~$D_1$ and~$D_2$, defined as
\begin{equation}
\boldsymbol D_1 \equiv -\k_1-\k_2 \;, \qquad {\rm and } \qquad \boldsymbol D_2 \equiv -\k_2-\k_3 \;.
\end{equation}
The full data vector that depends on all momenta and diagonals is very long and has very large dimensional covariance matrix. For these reasons the compressed trispectrum is sometimes used in practice. One such example is the integrated trispectrum, where in the estimator one averages over all~$D_1$ and~$D_2$ for the fixed momenta~$k_1,\ldots,k_4$. Using this estimator significantly reduces the length of the data vector and allows for a simple evaluation of the covariance matrix. The integrated trispectrum has been used both in real and redshift space, for comparison of perturbation theory predictions to simulations~\cite{Gualdi:2020eag,Gualdi:2022kwz}. 

In this paper we want to estimate the impact of the trispectrum on analyses on large scales and find the {\rm most optimal} possible improvement. For this reason we cannot use the simplified estimators such as integrated trispectrum, where some information is lost due to integration over the diagonals. This is particularly relevant for understanding the impact of the trispectrum on constraining all cubic bias parameters, where the only hope to do it comes from using the full shape keeping all dependence on different angles. Given this, we will use the full connected trispectrum and the following likelihood
\begin{equation}
    \ln\mathcal{L}_T=-\frac{1}{2}\sum_{\mathcal Q} \frac{(T^c_g(\mathcal Q,\theta)-\hat T^c_g(\mathcal Q))^2}{\text{Cov}_T^c(\mathcal Q)}\;,
\end{equation}
where the sum runs over all nonequivalent quadrilaterals~$\mathcal Q$ defined by the six sides~$\mathcal Q\equiv\{ k_1,k_2,k_3,k_4,D_1,D_2\}$. We are again using the Gaussian approximation, in which only the same shape quadrilaterals correlate. Note that, in order to avoid the double counting, the sides of the quadrilaterals are ordered as $k_4\leq k_3\leq k_2\leq k_1$, while the diagonals must satisfy the relations~$|k_1-k_2|\leq D_1\leq |k_1+k_2|$ and~$|k_2-k_3|\leq D_2\leq |k_2+k_3|$.

The diagonal elements of the connected trispectrum covariance matrix can be computed for the general~$\mathcal{Q}$ and we give the derivation in the Appendix~\ref{app:covariances}. They are given by
\begin{equation}
\text{Cov}_T(\mathcal Q) 
=\frac{(2\pi)^9 s_{1234}}{V V_{1234}} P_g(k_1,\theta_{\rm fid})P_g(k_2,\theta_{\rm fid})P_g(k_3,\theta_{\rm fid})P_g(k_4,\theta_{\rm fid})\;,
\end{equation}
where $s_{1234}=n!$ and $0\leq n\leq 4$ is the number of momenta~$k_1, \dots, k_4$ with the same magnitude. The nontrivial result is the volume of the quadrilateral shell $V_{1234}$, which reads
\begin{align}\label{eq:shell}
V_{1234} & = \frac83 \pi^2 \frac{k_1 k_2 k_3 k_4 D_1 D_2}{V_{\rm tetra}} \Delta k^6  \; .
\end{align}
In this expression~$V_{\rm tetra}$ is the volume of the tetrahedron~$\mathcal{Q}$ (for details, see Appendix~\ref{app:covariances}). This is a very simple formula, which allows for an easy evaluation of the trispectrum likelihood.

We notice that some of the considered tetrahedra have the six sides which lie on the same plane resulting in a degenerate shape and in a null volume. Since their number is very small ($\sim10^1$) compared to the total ($\sim 10^5$) we exclude them from our analysis.

\subsection{BAO wiggles}

\noindent In addition to the baseline analysis of the correlation functions up to a given~$k_{\rm max}$, we will also consider information on the amplitude of the linear density field in the broadening of the BAO wiggles. The likelihood in this case has the same form as for the power spectrum, except that the signal and the model contain the wiggly part of the galaxy power spectrum only 
\begin{equation}
\label{eq:pw}
     \ln\mathcal{L}_{\rm w}= -\frac{1}{2}\sum_{k=k_{\rm max}}^{\bar{k}} \frac{(P_g^{\rm w}(k,\theta) -\hat P_g^{\rm w} (k))^2}{\text{Cov}_P(k)}\;.
\end{equation} 
In our implementation we choose~$\bar{k}=0.3\; h/\mathrm{Mpc}$ since the wiggles are essentially completely damped beyond this scale. Note that the covariance matrix is the same as for the power spectrum. Clearly, the SNR on the amplitude~$\alpha$ is suppressed by the amplitude of the wiggly part of the power spectrum which is smaller than~$\mathcal O(10\%)$. 

\subsection{Theoretical error}
\noindent In addition to the data covariance, it is often useful in perturbation theory based analysis to consider theoretical error covariance. Theoretical error takes into account realistic uncertainties in the perturbative modeling which serves two purposes. One is to prevent the overfitting by choosing aggressive data cuts. The other is to prevent biases in cosmological inference. Details of the formalism and applications to galaxy clustering and lensing can be found in~\cite{Baldauf:2016sjb,Chudaykin:2020hbf,Moreira:2021imm,Aires:2024wze,Maraio:2024xjz}. 

The elements~$(ij)$ of the theoretical error covariance matrix have the following form
\begin{equation}
    (C_e)_{ij}=E_iE_j \rho_{ij}\;,
\end{equation}
where the index~$i$ labels a bin in the data vector, which can be a~$k$ bin or a bispectrum or trispectrum configuration. Note that indices on the right hand side are not summed over. The quantity~$E_i$ is the measure of the absolute error for the theoretical prediction in a given bin~$i$ and can be obtained from estimating the higher-order perturbation theory contributions neglected in the calculation. The matrix~$\rho_{ij}$ measures the correlation between the two different bins and it is crucial to ensure that one effectively marginalizes only over sufficiently smooth theoretical predictions~\cite{Baldauf:2016sjb}. For the power spectrum it is chosen to be
\begin{equation}
\rho_{ij}^{\rm P} = \exp\left [ -\frac 12 \frac{(k_i-k_j)^2}{\Delta k^2} \right] \;,
\end{equation}
where~$\Delta k$ is the parameters that determines the smoothness scale, over which the theoretical prediction vary very smoothly. In particular, we choose~$\Delta k=0.1\; h/$Mpc. If the two~$k$-bins are separated by more than~$\Delta k$ they are effectively uncorrelated and the theoretical prediction can vary within the theoretical uncertainty given by~$E_i$ and~$E_j$. If the bins are within~$\Delta k$, they are constrained to effectively fluctuate coherently. In the case of the bispectrum, the correlation matrix is a bit more complicated and takes into account the shape of the triangle as follows
\begin{equation}
\rho_{ij}^{\rm B} = \exp\left [ -\frac 12 \frac{(k_{i,1}-k_{j,1})^2}{\Delta k^2} \right] \exp\left [ -\frac 12 \frac{(k_{i,2}-k_{j,2})^2}{\Delta k^2} \right] \exp\left [ -\frac 12 \frac{(k_{i,3}-k_{j,3})^2}{\Delta k^2} \right] \;.
\end{equation}

In conclusion, the theoretical error covariance matrix has two parameters. One is the coherence length~$\Delta k$ and the other one is the estimate of the absolute theoretical error in perturbation theory. Following~\cite{Baldauf:2016sjb} we use the following estimates for the two-loop power spectrum 
\begin{equation}
\label{eq:theo_err_p}
E^{\rm P,2-loop}_i = D^4(z) P_g(k_i,z) \left( \frac{k_i}{0.45\; h/{\rm Mpc}} \right)^{3.3} \;,
\end{equation}
and the one-loop bispectrum
\begin{equation}
\label{eq:theo_err}
E^{\rm B,1-loop}_i = D^2(z) B_g(k_{i,1},k_{i,2},k_{i,3},z) \left( \frac{(k_{i,1}+k_{i,2}+k_{i,3})/3}{0.3\; h/{\rm Mpc}} \right)^{1.8} \;.
\end{equation} 
Note that requiring the theoretical error to be of the order of a given loop size is the other way to impose the perturbativity prior, where the total amplitude of fluctuations in the galaxy number density on large scales in always small. 

Once the theoretical error covariance matrix is computed, it can be straightforwardly included in the analysis by computing the Gaussian likelihood with the total covariance 
\begin{equation}
C_{ij} = C_{ij}^{\rm data} + (C_e)_{ij} \;,
\end{equation}
where~$C_{ij}^{\rm data}$ is the usual data covariance that contains cosmic variance and the noise. Note that the new total covariance matrix is not diagonal and one has to generalize the formulas for the likelihood above accordingly. Nevertheless, the likelihood is still Gaussian.

\subsection{Total likelihood}

\noindent Assuming the Gaussian likelihood (with or without the theoretical error) for the power spectrum, bispectrum, trispectrum and possibly BAO wiggles, we will combine them in our analysis in a simple way as follows
\begin{equation}
\label{eq:total_lik}
\ln \mathcal L = \ln \mathcal L_P + \ln \mathcal L_B + \ln \mathcal L_T + \ln \mathcal L_{\rm w} \;.
\end{equation}
At first sight, this assumption may seem too optimistic, since we neglect all cross-covariance terms that may be relevant. In what follows we justify this choice. 

We can start from the BAO wiggles. When we use the wiggly part of the power spectrum, we always use only scales with~$k>k_{\rm max}$. This separation of scales justifies a simple combination of wiggles with the rest of the data vector. 

For the correlation functions at~$k<k_{\rm max}$, the simple form of the total likelihood in Eq.~\eqref{eq:total_lik} is justified by arguments in perturbation theory. Since we are focusing on very large scales, the nonlinear contributions to the covariance and cross-covariance among different observables is small. In particular, these contributions are smaller than the effective error on cosmological parameters that comes from marginalization over many nuisance parameters~\cite{Takahashi:2009ty}. This has been explicitly checked and verified for the power spectrum and bispectrum analyses on relevant scales (e.g., \cite{Wadekar:2019rdu,Oddo:2019run,Martin:2011xt,Takahashi:2009bq,Takahashi:2009ty}. For example, the power spectrum analysis based on perturbation theory leads to identical results regardless of the choice of the covariance (Gaussian, non-linear or simulation-based), in a volume similar to the one we use in this paper~\cite{Tucci:2023bag,Wadekar:2020hax}. The same conclusion holds in the joint P+B analysis as shown in many works which studied this combination in various setups (see for example~\cite{Sefusatti:2006pa,Ivanov:2021kcd,Chan:2016ehg,Novell-Masot:2023gmj} and references therein). 

Finally, regarding the trispectrum, we expect that the similar arguments hold and that~Eq.~\eqref{eq:total_lik} is a very good approximation to the truth. Even if the analysis was optimistic, this is in line with our previous choices that maximize the possible information content of the trispectrum. One important point to make is that the trispectrum has also the cross-covariance with the power spectrum already at the Gaussian level. This can be in principle significant. However, in practice this is not the case. In Appendix~\ref{app:PTterms} we show that less than~$0.1\%$ of the total number of quadrilaterals correlates with the power spectrum at leading order, making such contributions to the total signal negligible. Therefore, they can be safely neglected in the analysis.

\subsection{Simple mode counting argument for the joint power spectrum and bispectrum analysis}

\noindent Before we move on to the verification of our model and approximations in simulations and the main analysis, let us first give a simple estimate of the results that we expect. 

The main advantage of using Eulerian perturbation theory is that the model is a simple polynomial in nuisance parameters and the amplitude of the linear density field. This means that the logarithm of the full likelihood in Eq.~\eqref{eq:total_lik} takes particularly simple form and one can easily estimate the precision with which different parameters can be measured. 

First of all, the highest signal to noise is in the power spectrum. On large scales the most dominant term is the linear theory contribution proportional to~$b_1^2\alpha^2$. Therefore, the galaxy power spectrum in real space measures the combination~$b_1\alpha$ very precisely. The relative error can be estimated as follows
\begin{equation}
\frac{\sigma(b_1\alpha)}{b_1\alpha} = \frac{1}{\sqrt{2N_{\rm pix.}}}\;,
\end{equation}
where the number of pixels in practice can be estimated as~$N_{\rm pix.}\approx Vk_{\rm max}^3/(2\pi)^3$. Note that this error is different from the error on the amplitude of the power spectrum, since the amplitude of the power spectrum is proportional to~$\alpha^2$. In practice, this is the most-well measured number in the real space analysis and we can essentially treat it as fixed. The degeneracy between the linear bias and~$\alpha$ is broken at the level of the power spectrum only very mildly by the loop contributions. However, given several free higher order bias parameters, the amplitude~$\alpha$ is essentially unconstrained by the power spectrum alone. 

Moving to higher order statistics, assuming~$b_1\alpha$ is fixed, the amplitude of the tree-level bispectrum is proportional to a single power of~$\alpha$. While the bispectrum also contains quadratic biases as free parameters, the angular dependence allows for degeneracy with~$\alpha$ to be partially broken. Therefore, we can assume that the error on the amplitude of the linear field from the joint power spectrum and bispectrum analysis is given by the bispectrum SNR. We can write
\begin{equation}
\label{eq:estimate_err_alpha}
\frac{\sigma(\alpha)}{\alpha} \approx {\rm few}\times \frac{1}{\Delta(k_{\rm max})\sqrt{N_{\rm pix.}}} \;,
\end{equation}
where we have used Eq.~\eqref{eq:snrest} and the numerical factor of a few depends on how much information is lost by marginalizing over~$b_2$ and~$b_{\mathcal G_2}$, which can be easily obtained using the Fisher matrix. 

Using~$V=(2\;{\rm Gpc}/h)^3$ and~$k_{\rm max}=0.1 \; h/$Mpc, we can estimate
\begin{equation}
\frac{\sigma(b_1\alpha)}{b_1\alpha} \approx 0.004\;, \qquad {\rm and} \qquad \frac{\sigma(\alpha)}{\alpha} \approx {\rm few} \times 0.02 \;.
\label{eq:est_errs}
\end{equation}
These numbers give an idea of how well the P+B analysis in real space should perform. In reality, the constraints are a bit weaker, due to the noise and nonlinearities we neglected in these arguments. However, we can already see that the baseline expectation for the error on~$\alpha$ is roughly~$\mathcal{O}(5-10\%)$, which agrees with the field level result of~\cite{Nguyen:2024yth}. Furthermore, the simple estimates above give us two additional important predictions. 

First, since there are no as strong degeneracies in the bispectrum as the~$b_1-\alpha$ degeneracy in the power spectrum, the trispectrum is not expected to significantly improve the errors. This is due to much smaller SNR in the trispectrum as well as existence of new cubic bias parameters that have to be measured as well. We expect~$\Delta^2(k_{\rm max}) \approx\mathcal{O}(10\%)$ improvements from inclusion of the trispectrum in the analysis. Second, in a P+B analysis we get a scaling of the error with~$k_{\rm max}$ which is given by
\begin{equation}
\label{eq:relative_err_alpha_pred}
\frac{\sigma(\alpha)}{\alpha} \sim \frac{k_{\rm max}^{-3/2}}{\Delta(k_{\rm max})} \;.
\end{equation}
Note that the typical amplitude of fluctuations~$\Delta(k_{\rm max})$ grows with~$k_{\rm max}$, which implies that the error on~$\alpha$ shrink faster than the naive~$k_{\rm kmax}^{-3/2}$ scaling. This makes sense, since the information comes from the bispectrum whose SNR grows as the field becomes more non-Gaussian on large scales, in addition to the simple mode counting. Note that the higher order~$n$-point functions would lead to different scaling with~$\Delta(k_{\rm max})$. This is yet another way to estimate where does the information come from, by carefully measuring the dependence of the inferred errors on~$k_{\rm max}$. 

In what follows, we will show that all three predictions (the size of the error bars, importance of the trispectrum and scaling with~$k_{\rm max}$) are confirmed in a real analysis.

\section{Validation on simulations}\label{sec:simul}

\noindent The goal of this section is to validate our model and assumptions using large-volume numerical simulations. Our results are not new, this has been already done in a large body of work on the power spectrum and bispectrum and comparison to large-volume simulations with a similar setup~\cite{Sefusatti:2006pa,Sefusatti:2010ee,Oddo:2019run,MoradinezhadDizgah:2020whw,Alkhanishvili:2021pvy,Eggemeier:2021cam,Oddo:2021iwq,Ivanov:2021kcd}. In particular, we use the set of 10 boxes of the PT challenge simulation suite~\cite{Nishimichi:2020tvu} (we refer the
reader to this reference for details on the simulations). With the total volume of 566 $({\rm Gpc}/h)^3$ at redshift~$z=0.6$, these simulations allow for very precise comparison of data and theory even on very large scales. We use the measured power spectrum and bispectrum of dark matter halos in real space, with number density and redshifts similar to the one analyzed in~\cite{Nguyen:2024yth}.

In particular, we want to test the following:
\begin{itemize}
    \item Eulerian perturbation theory predicts the power spectrum and the bispectrum on large scales well enough and leads to unbiased inference of~$\alpha$. 
    \item Large displacements on large scales do not produce any effect beyond the one captured by Eulerian perturbation theory. In particular, one does not have to do the infrared resummation. The difference between the infrared resummed and baseline model appears only on smaller scales, beyond $k_{\rm max}$ used in our analysis. 
    \item Typical deviation of the best-fit model and the data for~$k>k_{\rm max}$ is within the theoretical error estimate. This means that the realistic field of biased tracers has small fluctuations in agreement with perturbativity prior. Correspondingly, any combination of higher order bias and nonlinear correction has to be small on the scales considered, even though all bias coefficients may be of order~$\mathcal O(1)$. 
    \item The error bars on the amplitude~$\alpha$ are roughly given by the simple estimate in Eq.~\eqref{eq:estimate_err_alpha}. The error bars do not change significantly when the theoretical error is included in the analysis, indicating that the higher bias and loop corrections are not very relevant on very large scales used in the analysis. 
\end{itemize}

In order to do this, we will use the measurements from the full volume of 566 $({\rm Gpc}/h)^3$, but fit it with the covariance matrix rescaled to match the box of~8 $({\rm Gpc}/h)^3$ which has the same volume as in~\cite{Nguyen:2024yth}. We will also use the same two cuts,~$k_{\rm max}=0.1 \; h/{\rm Mpc}$ and~$k_{\rm max}=0.12 \; h/{\rm Mpc}$. We perform the joint power spectrum and bispectrum analysis using the theoretical model and likelihoods described above. We vary the following parameters
\begin{equation}
\theta  = \{\alpha, b_1, b_2, b_{\mathcal G_2}, b_{\Gamma_3}, c_s^2, c_1, c_2, c_3 \} \;,
\end{equation}
that include all the necessary bias parameters to compute the one-loop power spectrum and the tree-level bispectrum.
It is important to emphasize that we also make an appropriate binning of the model, integrating the theoretical prediction in each bin with the appropriate weight to take into account the number of Fourier modes for each momentum~$k$. This is important for the lowest~$k$-bins, but it does not significantly affect the final best fit and the error bars, since they are predominantly determined by the highest~$k$ modes in the analysis. 

\begin{figure}
    \centering
    \includegraphics[width=0.49\linewidth]{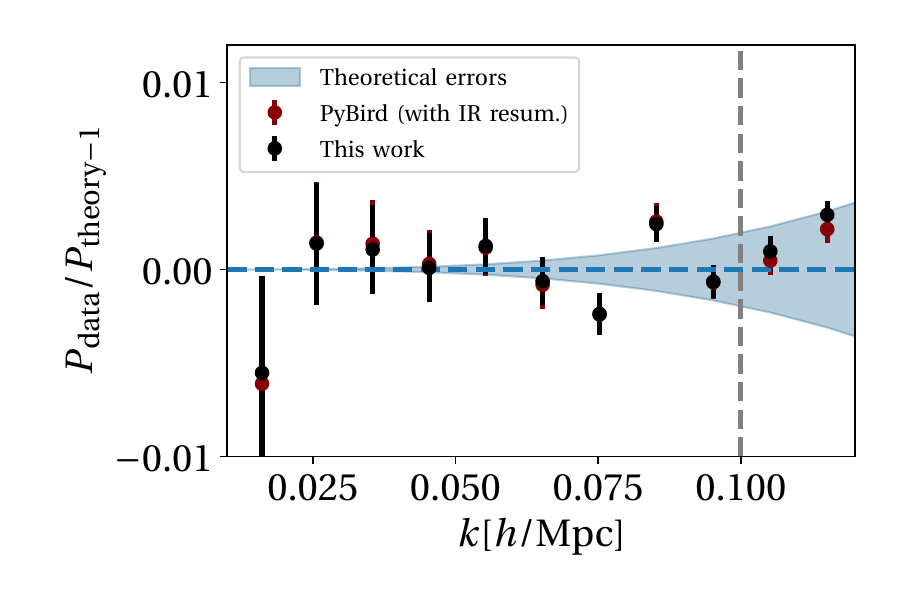}
     \includegraphics[width=0.49\linewidth]{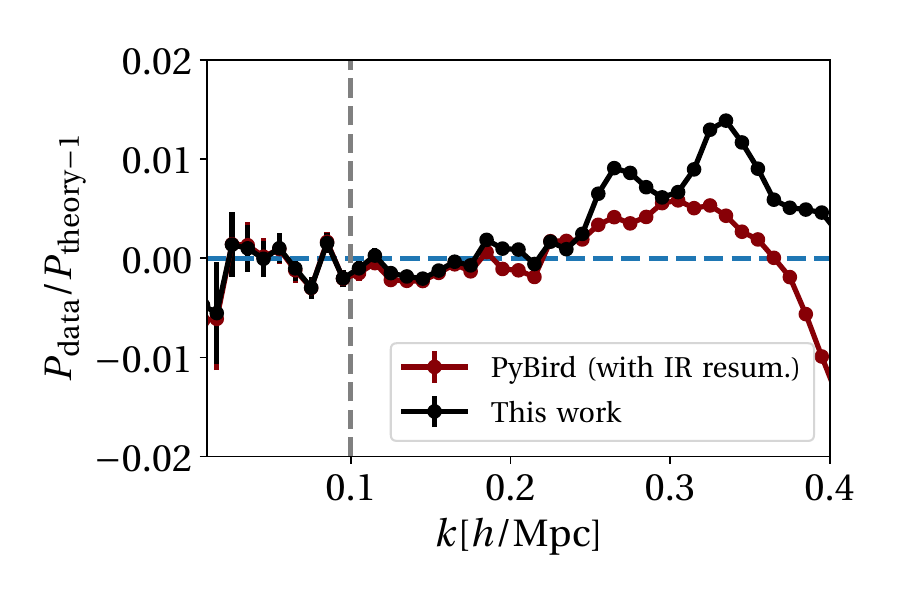}

    \caption{Comparison between the data and the best-fit model in a joint power spectrum and bispectrum analysis up to $k_{\rm max}=0.1\; h/$Mpc at~$z=0.6$ for the galaxy power spectrum. In the left panel the shaded region represents the 2-loop theoretical error of Eq.~\eqref{eq:theo_err_p}. An independent fit obtained using the \texttt{PyBird} code which accounts for the IR resummation is also shown. Statistical error bars on data correspond to the full PT Challenge volume. }
    \label{fig:PS_fit}
\end{figure}

\begin{figure}
    \centering
    \includegraphics[width=0.45\linewidth]{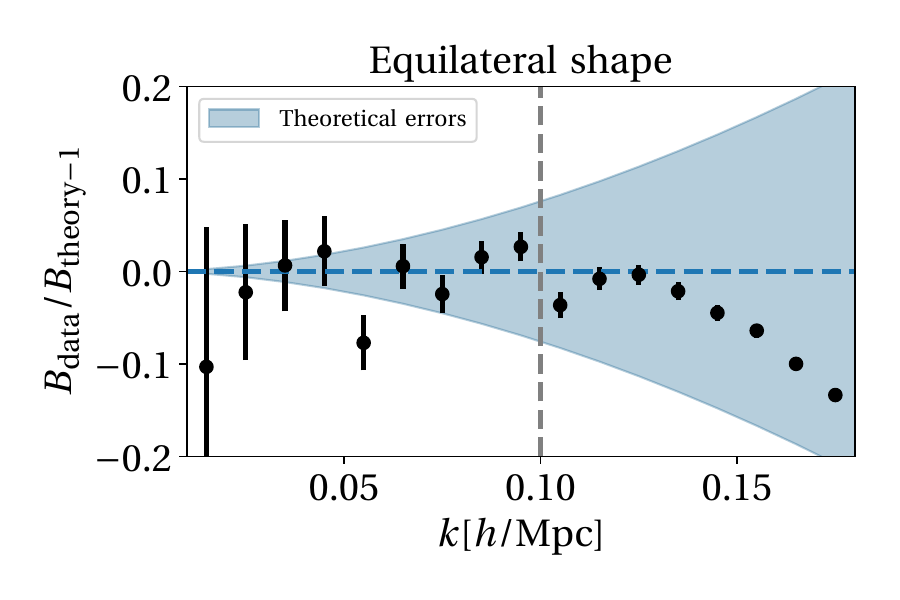}
    \includegraphics[width=0.45\linewidth]{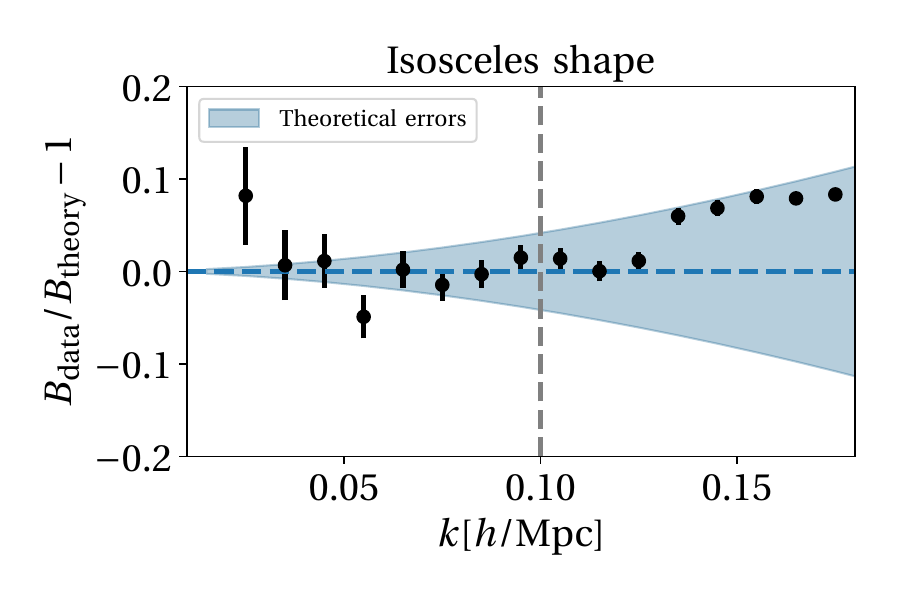}    \caption{Comparison between the data and the best-fit model in a joint power spectrum and bispectrum analysis up to $k_{\rm max}=0.1\; h/$Mpc at~$z=0.6$ for the equilateral and isosceles (i.e., $k_1=0.015\;h/\mathrm{Mpc}
    ,k_2=k_3$) shapes of galaxy bispectrum. Statistical error bars on data correspond to the full PT Challenge volume. The shaded region represents the 1-loop theoretical error of Eq.~\eqref{eq:theo_err}. } 
    \label{fig:bisp_fit}
\end{figure}

\begin{figure}
    \centering
    \includegraphics[width=0.45\linewidth]{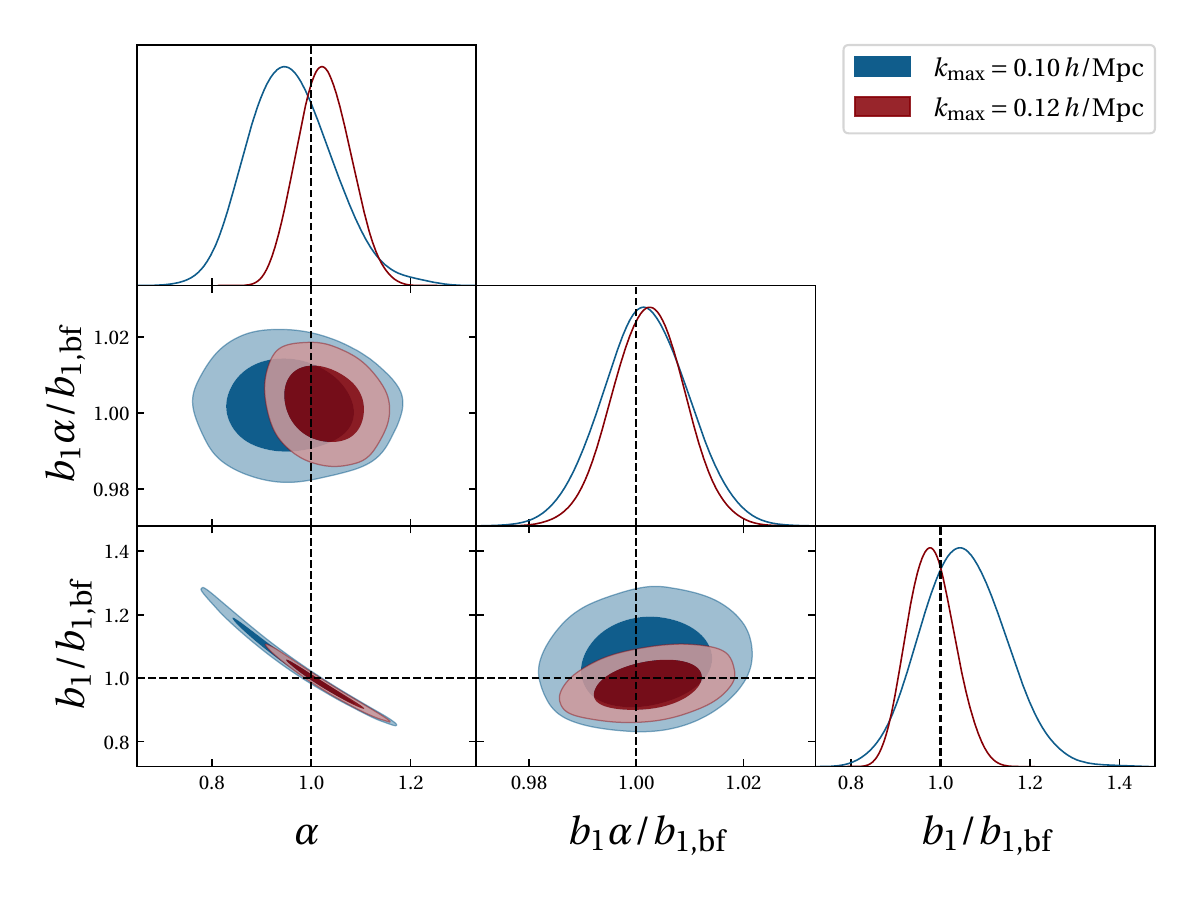}
    \caption{P+B analysis of PT challenge data. The covariance used in the analysis is computed using the reduced volume~$V=(2\; {\rm Gpc}/h)^3$.}
    \label{fig:PTchallenge}
\end{figure}

The main results are shown in~Fig.~\ref{fig:PS_fit},~Fig.~\ref{fig:bisp_fit} and Fig.~\ref{fig:PTchallenge} and summarized in Tab.~\ref{tab:PT_PB_kmax}. The first two figures show the residual plots for the power spectrum and the bispectrum obtained from the joint analysis up to~$k_{\rm max}=0.1 \; h/$Mpc. In all plots the statistical error bars are shown for the full PT Challenge volume, which is roughly 70 times larger than the volume of~$V=8\,({\rm Gpc}/h)^3$ that we are interested in. For this reason all error bars are extremely small. For instance, the relative statistical errors in each~$k$-bin in the power spectrum are much smaller than~$1\%$. This exquisite precision allows us to do the stringent test of the model. 

Let us first focus on the power spectrum residuals shown in Fig.~\ref{fig:PS_fit}. The left panel shows the residuals together with the estimate of the 2-loop theoretical error. We can see that the best fit reproduces the data well within the theoretical uncertainties of perturbation theory. The two points with statistically significant deviation (2-3 sigma) are likely a random fluctuation, since they cannot be fitted by a sufficiently smooth curve. To verify this, we also fit the same data with the nonlinear \texttt{PyBird} code~\cite{DAmico:2020kxu} and find almost identical results\footnote{Equivalent approaches for large scale structure correlators have been implemented in other publicly available codes, such as \texttt{CLASS-PT}~\cite{Chudaykin:2020aoj}, \texttt{velocileptors}~\cite{Chen:2020zjt} and \texttt{CLASS-1loop}~\cite{Linde:2024uzr}.}. Importantly, the higher~$k$-bins beyond~$k_{\rm max}=0.1 \;h/$Mpc used in the analysis, lay within the estimate of the 2-loop theoretical error, indicating that the total galaxy field indeed follows expectations from perturbation theory.   
It is instructive to do the extrapolation of the best-fit model with~$k_{\rm max}=0.1 \;h/$Mpc to smaller scales. This is shown on the right panel of Fig.~\ref{fig:PS_fit}. The residuals remain within the theoretical error which grows very fast with~$k$. More importantly, one can see that using the Eulerian model which is the baseline for our analysis is insufficient on small scales to capture the BAO wiggles appropriately. However, the residual BAO wiggles appear only for~$k\gtrsim0.2\;h/$Mpc, which is never used in our baseline analysis. This is an important confirmation that large displacements on large scales are captured well by Eulerian perturbation theory. 

Similar residual plots for the bispectrum are shown in Fig.~\ref{fig:bisp_fit} for two different triangle configurations. One can see that the model fits the data well and that the residuals for~$k>k_{\rm max}$ are within the estimate of the one-loop theoretical error. 

It is important to emphasize once again that the statistical errors in these two figures are roughly 8 times smaller than the case of interest corresponding to the volume~$V=8\;({\rm Gpc}/h)^3$. This implies that for our main analysis, the theoretical model presented above is perfectly adequate to capture all relevant nonlinearities and it should lead to unbiased answers. This can be verified using the covariance matrix rescaled to the volume of the baseline analysis, still fitting the PT Challenge data. The corresponding triangle plots for $\alpha,b_1,b_1\alpha$ in this analysis for two different choices of~$k_{\rm max}$ are shown in Fig.~\ref{fig:PTchallenge}. One can see that the inference of the amplitude and~$b_1$ is unbiased, indicating that the simple model presented in Sec.~\ref{sec:models} guarantees a sufficient level of accuracy in describing the galaxy overdensity field on large scales and for the volume considered. 

\begin{table}
    \centering
    \begin{tabular}{|c|c|} \hline  
         & P+B\\ \hline 
 \multicolumn{2}{|c|}{$k_{\rm max}=0.1\; h/\mathrm{Mpc}$}\\ \hline  
         $\alpha$&  0.960$\pm$0.085\\ \hline  
         $b_1/b_{1,\mathrm{fid}}$&  1.051$\pm$0.094\\ \hline  
 \multicolumn{2}{|c|}{$k_{\rm max}=0.12\; h/\mathrm{Mpc}$}\\ \hline
 $\alpha$& 1.026$\pm$0.052\\ \hline 
 $b_1/b_{1,\mathrm{fid}}$& 0.979$\pm$0.051\\ \hline 
    \end{tabular}
    \caption{$1\sigma$ constraints and best fit parameter for the PT Challenge P+B analysis using the covariance with the volume of~$V=(2\;{\rm Gpc}/h)^2$.}
    \label{tab:PT_PB_kmax}
\end{table}

It is also interesting to look at the behavior of errors in Fig.~\ref{fig:PTchallenge} and Tab.~\ref{tab:PT_PB_kmax} as a function of~$k_{\rm max}$. We can see the improvement on the combination~$b_1\alpha$ is smaller than the mode counting expectation. This is due to the fact that marginalization over higher order biases effectively introduces an error which makes the constraints weaker. On the other hand, the errors on~$b_1$ and~$\alpha$ improve significantly with the increasing~$k_{\rm max}$. This is in line with the expectation from the information content of the bispectrum, where relative errors improve faster than~$k_{\rm max}^{-3/2}$ in agreement with Eq.~\eqref{eq:relative_err_alpha_pred}. In conclusion, the joint P+B analysis of the PT Challenge simulation boxes shows that the theoretical model presented in the previous section fits the data well and leads to errors in agreement with basic arguments based on perturbation theory. 

\begin{figure}
    \centering
    \includegraphics[width=0.45\linewidth]{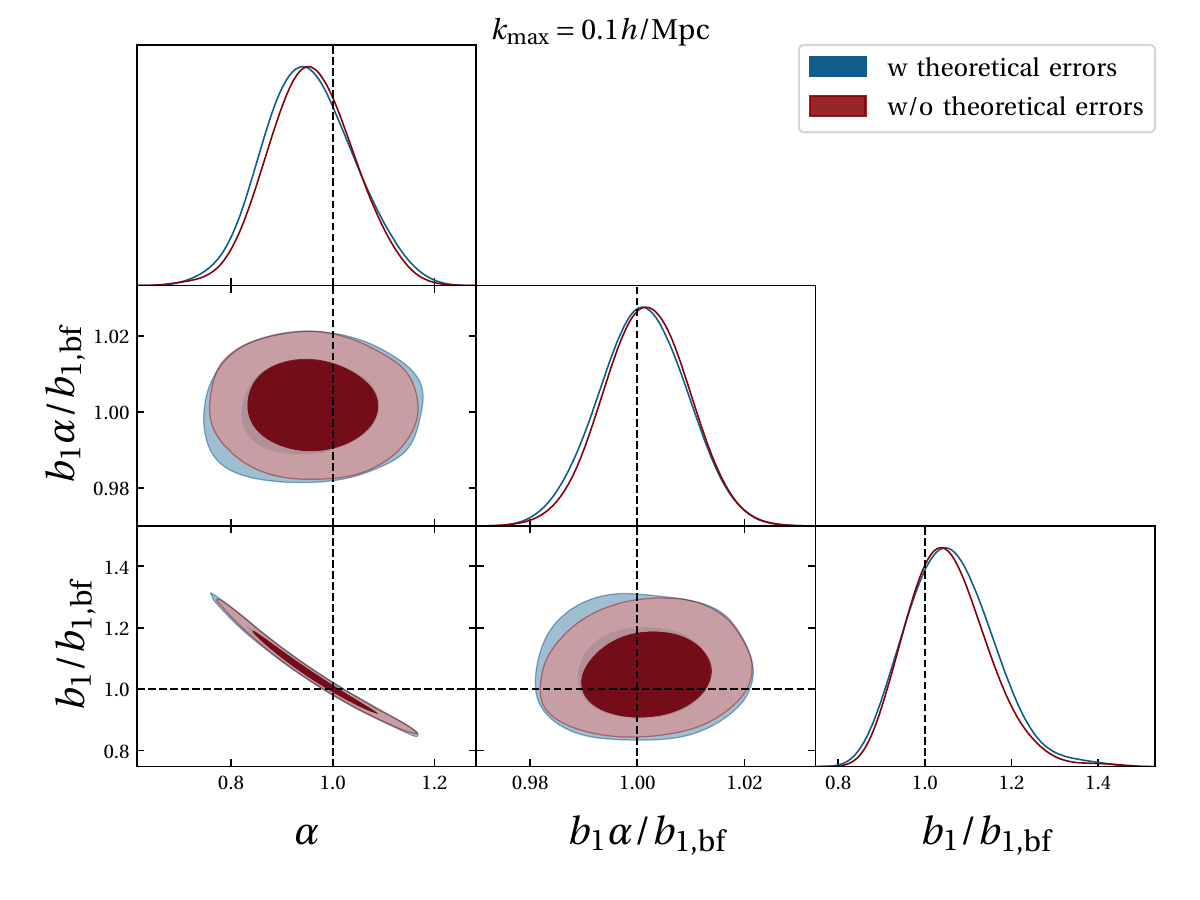}
    \includegraphics[width=0.45\linewidth]{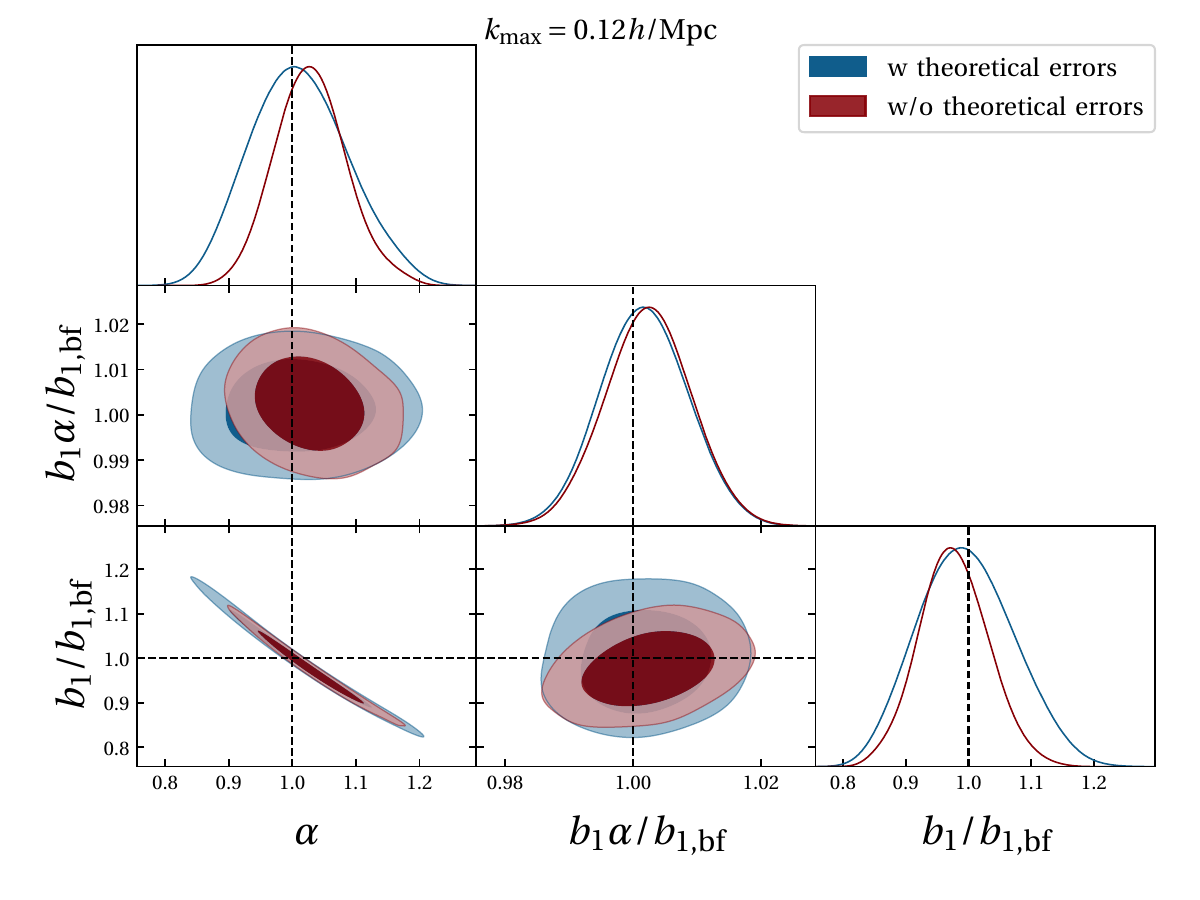}
    \caption{P+B analysis of the PT challenge data with and without theoretical errors for (\textit{Left}) $k_{\rm max}=0.1\; h/\mathrm{Mpc}$ and (\textit{Right}) $k_{\rm max}=0.12\; h/\mathrm{Mpc}$.  The covariance used in the analysis is computed using the reduced volume~$V=(2\; {\rm Gpc}/h)^3$.}
    \label{fig:theo_err}
\end{figure}

It is interesting to ask how this picture changes when the theoretical error is included in the analysis. As we discussed, the theoretical error is an additional correlated covariance whose purpose is to mimic marginalizing over neglected higher order corrections in the perturbation theory calculations. We repeat the same analysis of the PT Challenge but with the 2-loop power spectrum and 1-loop bispectrum theoretical errors included in the covariance. The results are shown in Fig.~\ref{fig:theo_err}. The triangle plots for $\alpha,b_1,b_1\alpha$ are shown with and without theoretical errors in the covariance, for two different values of~$k_{\rm max}$. The inclusion of the theoretical error makes the constraints on~$\alpha$ worse by~$10\%$ and~$30\%$ at~$k_{\rm max}=0.1\;h/\mathrm{Mpc}$ and~$k_{\rm max}=0.12\;h/\mathrm{Mpc}$ respectively. Such results are expected. On larger scales, the impact of theoretical error is much smaller, given that the model is more precise. Nevertheless, the inclusion of the theoretical error does not change the error bars dramatically. In particular, they cannot explain the difference of roughly factor of 3 between the result we find here and the SBI results for the joint P+B analysis in~\cite{Nguyen:2024yth}. We will come back to this point in Sec.~\ref{sec:discussion}.


\section{Results}\label{sec:results}

\noindent After extensive tests of the simple EPT model on large scales using large-volume PT Challenge simulations and verifying that it is adequate for the joint P+B analysis, we are finally ready to present the main results. Our goals are to estimate the error bars of the conventional P+B analysis in a box of the volume~$V=8\;({\rm Gpc}/h)^3$ for tracers similar to the one studied in~\cite{Nguyen:2024yth} and~\cite{Beyond-2pt:2024mqz} and estimate the possible improvements coming from including the galaxy trispectrum in the analysis.

We will not analyze the real simulation data due to unavailability of the trispectrum that is an important part of our results. Instead, using the baseline model (for which we know to match the reality on large scales) we generate the mock data vector containing P, B and T and the wiggly part of the power spectrum for~$k>k_{\rm max}$, and perform the analysis of this data vector. In order to do that, for simplicity, we use the same baseline cosmology as the PT Challenge which is very close to standard~$\Lambda$CDM universe. We also have to specify some fiducial values for the following set of parameters:
\begin{equation}
\theta = \{\alpha, b_1, b_2, b_{\mathcal G_2}, b_3, b_{\mathcal G_3}, b_{\mathcal G_2 \delta}, b_{\Gamma_3}, c_s^2, c_{\rm w}^2, c_1, c_2, c_3 \} \;.
\end{equation}
Obviously~$\alpha_{\mathrm{fid}}=1$. For other parameters we use two different sets of fiducial values depending on the tracer.

First, we chose the linear bias~$b_{1,\mathrm{fid}}=1.7$. We also set the noise to be given by the number density of the dark matter halos in real space equal to~$\bar n=1.3\times10^{-3} \; ($Mpc$/h)^3$ at fixed redshift~$z=0.5$. The linear bias, the noise and redshift are chosen to be similar to the SNG sample studied in~\cite{Nguyen:2024yth}. The fiducial values of other parameters are far less important, and we verify that explicitly in our analysis. Since the sample is dense, the nonlinear bias parameters are small. We choose the following values: $b_{2,{\rm fid}}=b_{\mathcal{G}_{2,{\rm fid}}}=b_{{\Gamma_3,{\rm fid}}}=b_3=b_{{\mathcal{G}_3,{\rm fid}}}=b_{{\delta\mathcal{G}_2,{\rm fid}}}=0.1$. The fiducial counterterms are given by~$c_{s,{\rm fid}}^2=1, c_{{\rm w},{\rm fid}}^2=0$ and we choose that all fiducial noise free parameters are equal to 1,~$c_{1,{\rm fid}}=c_{2,{\rm fid}}=c_{3,{\rm fid}}=1$. 
When running the MCMC chains we use the following priors: 
\begin{equation}
\label{eq:priorsforsample1}
    \begin{gathered}
        \mathcal{P}(\alpha) = \mathcal{U}(0.5, 1.5), \\
\mathcal{P}(b_1) = \mathcal{N}(1.0, 5.0), \\
\mathcal{P}(b_2) = \mathcal{P}(b_{\mathcal{G}_2}) = \mathcal{P}(b_{\Gamma_3}) =\mathcal{P}(b_{3}) =\mathcal{P}(b_{\mathcal{G}_3}) =\mathcal{P}(b_{\delta\mathcal{G}_2})=\mathcal{N}(0., 1.0),\\
\mathcal{P}(c_s^2) = \mathcal{N}(0, 5.0),\\
\mathcal{P}(c_{\mathrm{w}}^2) = \mathcal{N}(0, 1.0),\\
\mathcal{P}(c_1) = \mathcal{P}(c_2)  = \mathcal{P}(c_3)  =\mathcal{N}(1, 1)\;.
    \end{gathered}
\end{equation}
The variance of each parameter is chosen to match the choice in~\cite{Nguyen:2024yth}. 

The second set of fiducial nuisance parameters is chosen to describe the sample at~$z=1$ used in the beyond 2-point function challenge paper~\cite{Beyond-2pt:2024mqz}. From~\cite{Beyond-2pt:2024mqz} we estimate that~$b_1\approx 2.5$ and~$\bar{n}\sim4.5\times10^{-4} (h/\mathrm{Mpc})^3$. For the fiducial values of the higher order bias parameters we use the measurements for a similar linear bias in~\cite{Lazeyras:2015lgp,Lazeyras:2017hxw}. This is not exact, but it provides a good estimate of the nonlinear biases which for less dense samples can be large. Labeling with a tilde the biases in~\cite{Lazeyras:2015lgp,Lazeyras:2017hxw}, the rotation to our basis is
\begin{equation}
    b_1 = \tilde b_1\;, \quad b_{\mathcal G_2} = \tilde b_{K^2}\, \quad b_2 = \tilde b_2 + \frac43 \tilde b_{K^2} \;,
\end{equation}
for linear and quadratic bias and 
\begin{equation}
b_{\Gamma_3} = \tilde b_{\rm td}\;, \quad b_{\mathcal G_3} = - \tilde b_{K^3} \;, \quad b_{\mathcal G_2\delta} = \tilde b_{\delta K^2} + \frac 12 \tilde b_{K^3} - \frac8{21}\tilde b_{\rm td} \;, \quad b_3 = \tilde b_3 + 4 \tilde b_{\delta K^2} + \frac43 \tilde b_{K^3} \;,
\end{equation}
for cubic biases. Relaying on the measurements in~\cite{Lazeyras:2015lgp,Lazeyras:2017hxw} we estimate
\begin{equation}
b_1 \approx 2.5\;, \quad b_2 \approx 0.2 \;, \quad b_{\mathcal G_2} \approx -0.6 \;, \quad b_{\Gamma_3} \approx 1.5 \;, \quad b_{\mathcal G_2\delta} \approx 0.5 \;, \quad b_{\mathcal G_3} \approx -1 \;, \quad b_3 \approx -2 \;.
\end{equation}
The other parameters are the same as before. In particular,~$c_{s,{\rm fid}}^2=c_{1,{\rm fid}}=c_{2,{\rm fid}}=c_{3,{\rm fid}}=1$. We also use the same priors as in Eq.~\eqref{eq:priorsforsample1}.

We checked that a different choice of fiducial values of the nonlinear bias parameters  in both cases does not significantly impact our results, beyond small~$10\%$ corrections. More generally, all our results should be correct only up to~$10-20\%$, given several approximations that we use in the likelihood and the uncertainty in the fiducial parameters. This is acceptable, since we do not aim to have correct errors to better precision. We want to answer the question whether the error bars in FBI differ from P+B very significantly (by more than a factor of 2). Any mistake of the order of~$10\%$ is not very important for answering this question. 

\subsection{Joint power spectrum and bispectrum analysis}

\noindent We begin by presenting the analysis performed on the mock data for a sample in~\cite{Nguyen:2024yth} generated as described above, using the power spectrum and the bispectrum. The results are shown in Fig.~\ref{fig:PBT01} and Table~\ref{tab:PBT01}. There are three important points to make regarding the joint P+B analysis:
\begin{itemize}
    \item First, the error bars are in rough agreement with the mode counting estimate in Eq.~\eqref{eq:estimate_err_alpha}. This is the confirmation that the main degeneracy~$\alpha-b_1$ is broken by the bispectrum and that the remaining degeneracies of the amplitude~$\alpha$ with the nonlinear biases~$b_2$ and~$b_{\mathcal G_2}$ are much milder (even though they degrade the errors by a factor of a few compared to the ideal case where all biases are known). 
    \item Second, the error bars from a joint P+B analysis are much closer to the FBI results of~\cite{Nguyen:2024yth} and roughly a factor of~3 smaller than the SBI errors in the same paper. This is a very interesting result and we will discuss it in more detail in the next section. 
    \item Finally, the error bars follow the estimate for the scaling with~$k_{\rm max}$ given in~Eq.~\eqref{eq:relative_err_alpha_pred}, providing another confirmation for our estimates. In particular, going from~$k_{\rm max} = 0.1\; h/\mathrm{Mpc}$ to~$k_{\rm max} = 0.12\; h/\mathrm{Mpc}$, we expect roughly a~$30\%$ improvement, in agreement with our analysis on the mock data. 
\end{itemize}

Having obtained the error bars from the joint analysis of power spectrum and bispectrum, we proceed to investigate additional possible sources of information on top of this baseline analysis.

\subsection{Adding the trispectrum}

\noindent The three-level four-point function is the first higher order statistic that can be consistently constructed when the galaxy overdensity field is evolved up to third order in perturbation theory. In fact, the trispectrum completes the set of correlation functions that can be reliably predicted with the cubic model. In this section we present the joint P+B+T analysis and investigate the impact of the trispectrum on constraining the amplitude~$\alpha$. 

\begin{figure}
    \centering
    \includegraphics[width=0.45\linewidth]{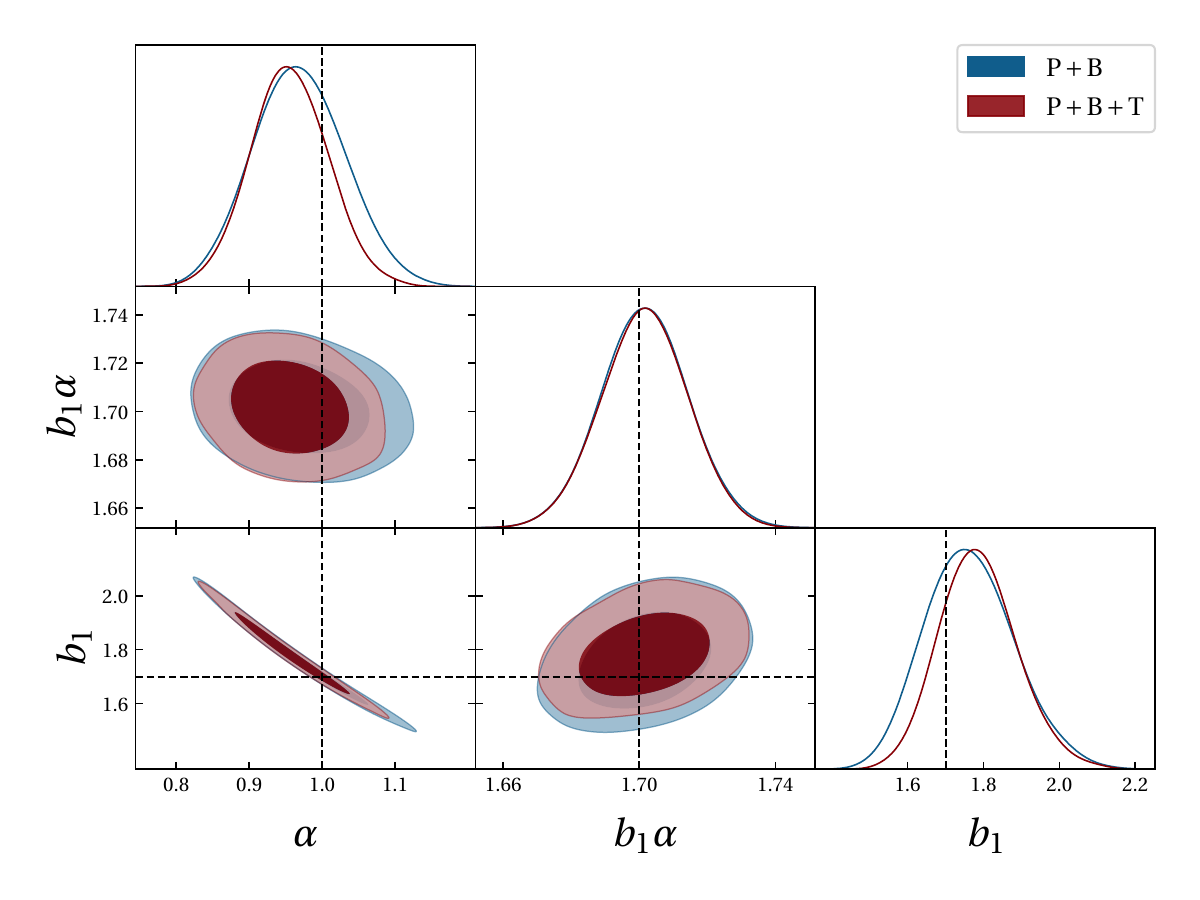}
    \includegraphics[width=0.45\linewidth]{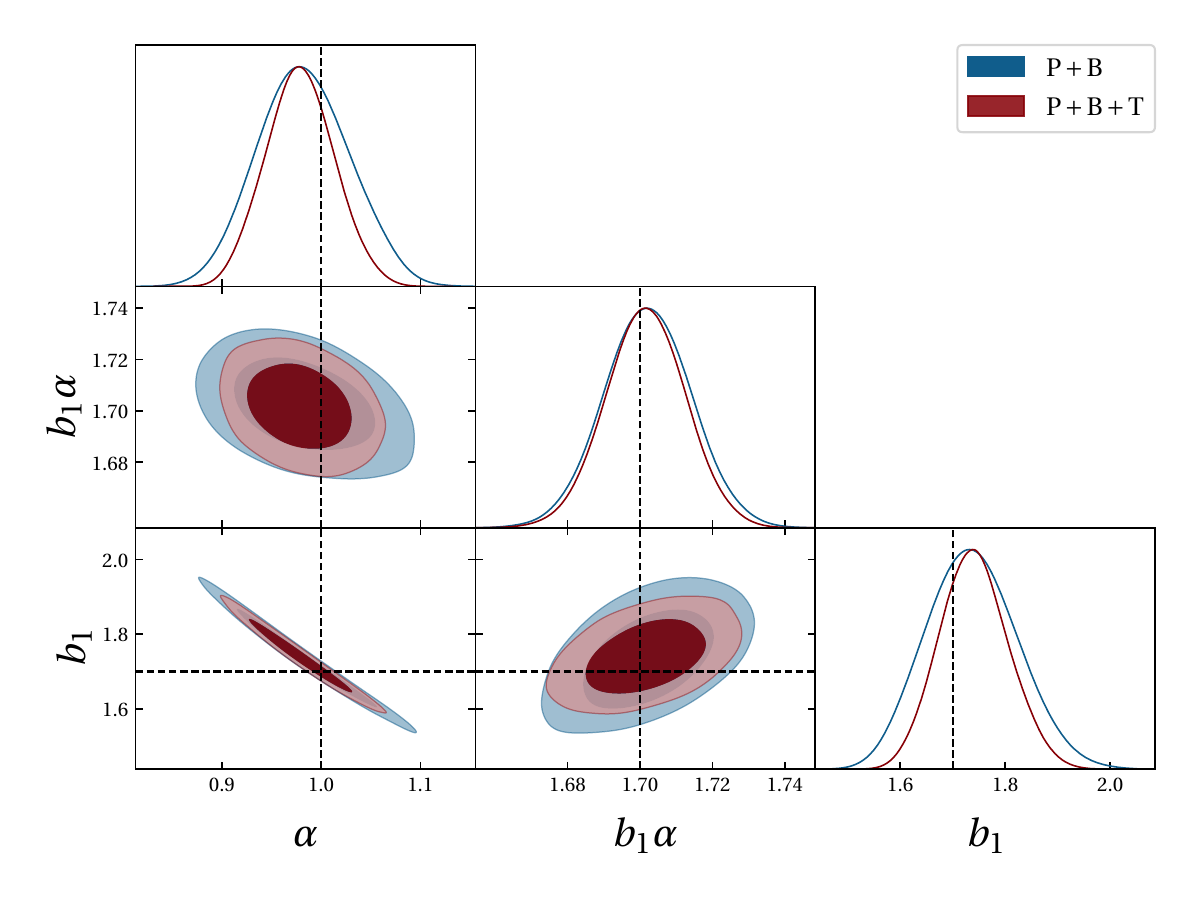}
    
    \caption{Comparison between P+B analysis and P+B+T using mock data for $k_{\rm max}=0.1\; h/$Mpc (\textit{Left}) and $k_{\rm max} = 0.12\; h$/Mpc (\textit{Right}).}
    \label{fig:PBT01}
\end{figure}

The results of the joint P+B+T analysis are shown in Fig.~\ref{fig:PBT01} and Table~\ref{tab:PBT01}. The two main findings are: 
\begin{itemize}
    \item Adding the trispectrum to the P+B analysis reduces the error bars on the amplitude~$\alpha$ by~$\sim20\%$ and~$\sim30\%$ at~$k_{\mathrm{max}}=0.1\, h/\mathrm{Mpc}$ and~$k_{\mathrm{max}}=0.12\; h/\mathrm{Mpc}$ respectively. This is an improvement in agreement with our SNR estimates. It indicates that there are no further major degeneracy breaking, which is expected given that there are several new free parameters in the trispectrum model. Nevertheless, even a~$\sim20\%$ reduction of error bars is interesting and it is important to see if the impact of the trispectrum is the same in a more realistic setup (e.g.~including the full nonlinear covariance or doing the analysis in redshift space).
    \item The error bars on~$\alpha$ at redshift~$z=0.5$ are 0.054 and 0.035 for $k_{\rm max} = 0.10\; h/\mathrm{Mpc}$ and $k_{\rm max} = 0.12\; h/\mathrm{Mpc}$ respectively. This is in a good agreement with the FBI analysis of~\cite{Nguyen:2024yth} (given our uncertainty in the results of~$10-20\%$). Such result is not obvious, and it is rather nontrivial. The two approaches are using different perturbation theory models and measure~$\alpha$ in two different ways. Therefore, assuming the absence of accidental cancellations, we interpret this agreement as a strong indication that the full complexity of the field level analysis on large scales can be captured by the leading~$n$-point functions computed in the simple Eulerian model, in agreement with expectations based on perturbation theory. 
\end{itemize}
.

\begin{table}
    \centering
    \begin{tabular}{|c|c|c|} \hline  
         &  P+B& P+B+T\\ \hline 
 \multicolumn{3}{|c|}{$k_{\rm max}=0.1\; h/\mathrm{Mpc}$}\\ \hline  
         $\alpha$&  0.968$\pm$0.064& 0.966$\pm$0.054\\ \hline  
         $b_1$&  1.764$\pm$0.118& 1.767$\pm$0.100\\ \hline  
 \multicolumn{3}{|c|}{$k_{\rm max}=0.12\; h/\mathrm{Mpc}$}\\ \hline
 $\alpha$& 0.981$\pm$0.046&0.979$\pm$0.035\\ \hline 
 $b_1$& 1.736$\pm$0.086&1.741$\pm$0.065\\ \hline 
    \end{tabular}
    \caption{$1\sigma$ constraints and best fit parameter for P+B and P+B+T analyses using mock data.}
    \label{tab:PBT01}
\end{table}

\begin{figure}
    \centering
      \includegraphics[width=0.5\linewidth]{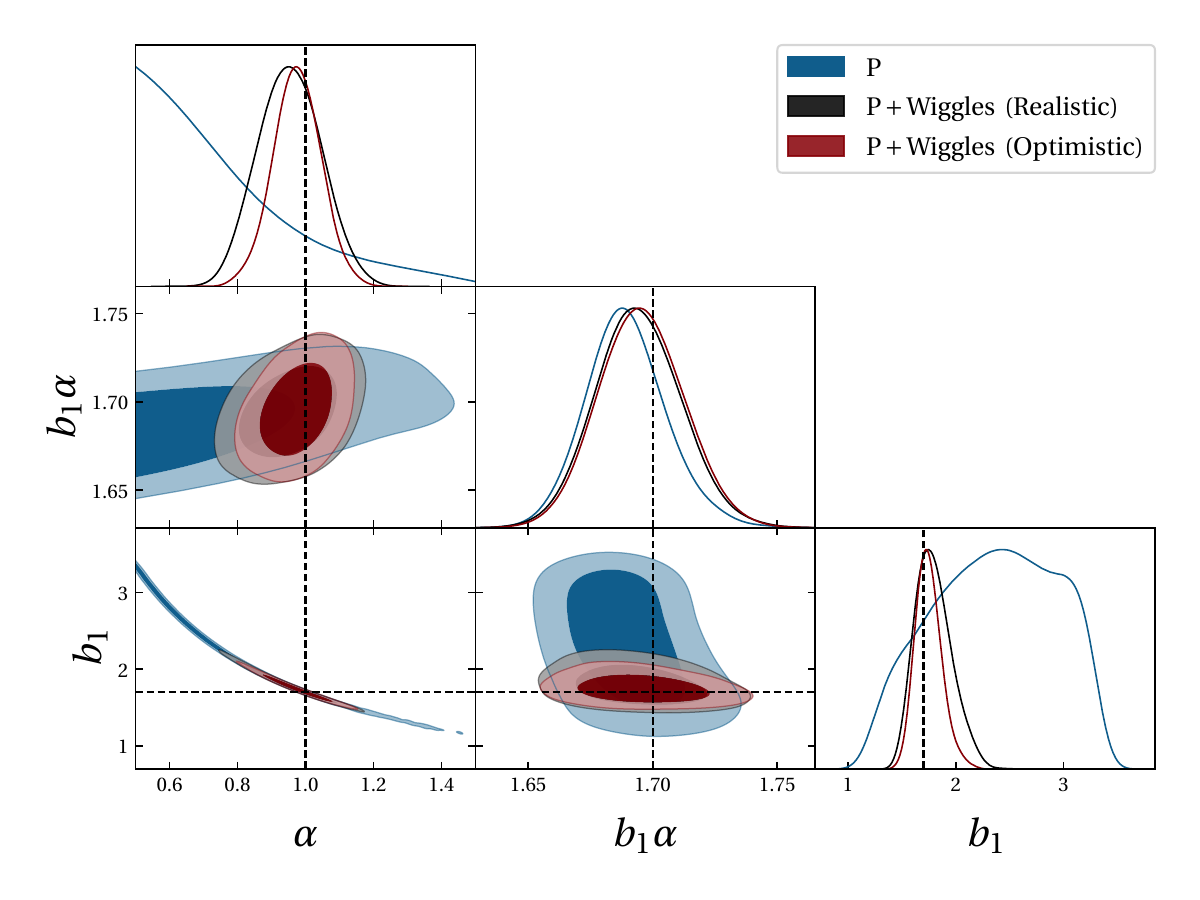}
   
    \caption{Comparison between constraints obtained using only the broadband fitted until $k_{\mathrm{max}}=0.1\; h/\mathrm{Mpc}$ and when adding also wiggles (with $k_\mathrm{{max,w}}=0.3\; h/\mathrm{Mpc}$).}  
        \label{fig:wiggles}
\end{figure}

\subsection{Adding the BAO wiggles}

\noindent As explained in Sec.~\ref{sec:likelihood} the broadening of the BAO peak can help to measure the amplitude~$\alpha$. This can improve the power spectrum only analysis where~$\alpha$ is measured only from the loop effects. The improvement is due to the fact that the damping factor~$\Sigma_\Lambda^2$ is proportional to~$\alpha$ only (and not to the combination $b_1 \alpha$) such that the amplitude of fluctuations can be estimated from the broadening the BAO peak. This effect is only due to the displacements and can be used to estimate how much information can be in principle recovered from the bulk flows which do not come with new free parameters. 

As previously discussed, we want to consider two limiting situations. In the first case (that we indicate as \textit{realistic})~$\Lambda$ is chosen to include only large-scale displacements (i.e., $\Lambda=0.2\; h/\mathrm{Mpc}$ as in~\cite{Chudaykin:2020aoj}) and one introduces a new free parameter $c^2_{\rm w}$ which plays the role of a counterterm absorbing all possible higher-order contributions to the wiggly power spectrum. The second case instead represents the best-case scenario (\textit{optimistic}) in which one sends~$\Lambda\rightarrow+\infty$ and sets $c^2_{\rm w}=0$. In this case, one assumes a perfect knowledge of all non-linear effects induced by large bulk flows and the full broadening of the BAO peak can be described without new parameters.

In Fig.~\ref{fig:wiggles} and~\ref{tab:wiggles} we show the constraints on~$b_1$ and~$\alpha$ obtained for the power spectrum up to~$k_{\rm max} = 0.1\; h/{\rm Mpc}$ with and without adding the information from the wiggles with~$k_{\rm max}<k\leq\bar{k} = 0.3\;h/{\rm Mpc}$. We present the constraints for the realistic and optimistic scenario. The main findings are:
\begin{itemize}
    \item The information on the amplitude~$\alpha$ in the wiggles is significantly higher than in the power spectrum at~$k<k_{\rm max}$. Even though the wiggles have small amplitude ($\sim10\%$ of the power spectrum), given the robustness of the BAO broadening, the constraints from the large displacements are stronger than the one-loop power spectrum. 
    \item Even though adding the wiggles improves the results, they are still not as good as adding the bispectrum on large scales. In principle, it is not obvious that this had to be the case. We conclude that in a~$\Lambda$CDM-like cosmology, in order to get optimal constraints, one has to include the higher-order~$n$-point functions. 
    \item The results do not change dramatically going from optimistic to realistic scenario. This implies that the constraints from the BAO wiggles are rather robust. Note that this information is automatically included in the full-shape analyses of the power spectrum that use~$k_{\rm max}\sim0.3\; h/{\rm Mpc}$. 
\end{itemize}

\begin{table}
    \centering
    \begin{tabular}{|c|c|c|c|} \hline 
         &   P&  P+Wiggles (realistic)& P+Wiggles (optimistic)\\ \hline 
         $\alpha$&   0.743$\pm$0.185&  0.95$\pm$0.099& 0.975$\pm$0.072\\ \hline 
         $b_1$&   2.402$\pm$0.535&  1.801$\pm$0.191& 1.748$\pm$0.124\\ \hline
    \end{tabular}
    \caption{$1\sigma$ constraints and best fit parameter for broadband analysis of the power spectrum and when including also the wiggles in the realistic and in the optimistic scenario.}
    \label{tab:wiggles}
\end{table}

\begin{figure}
    \centering
    \includegraphics[width=0.45\linewidth]{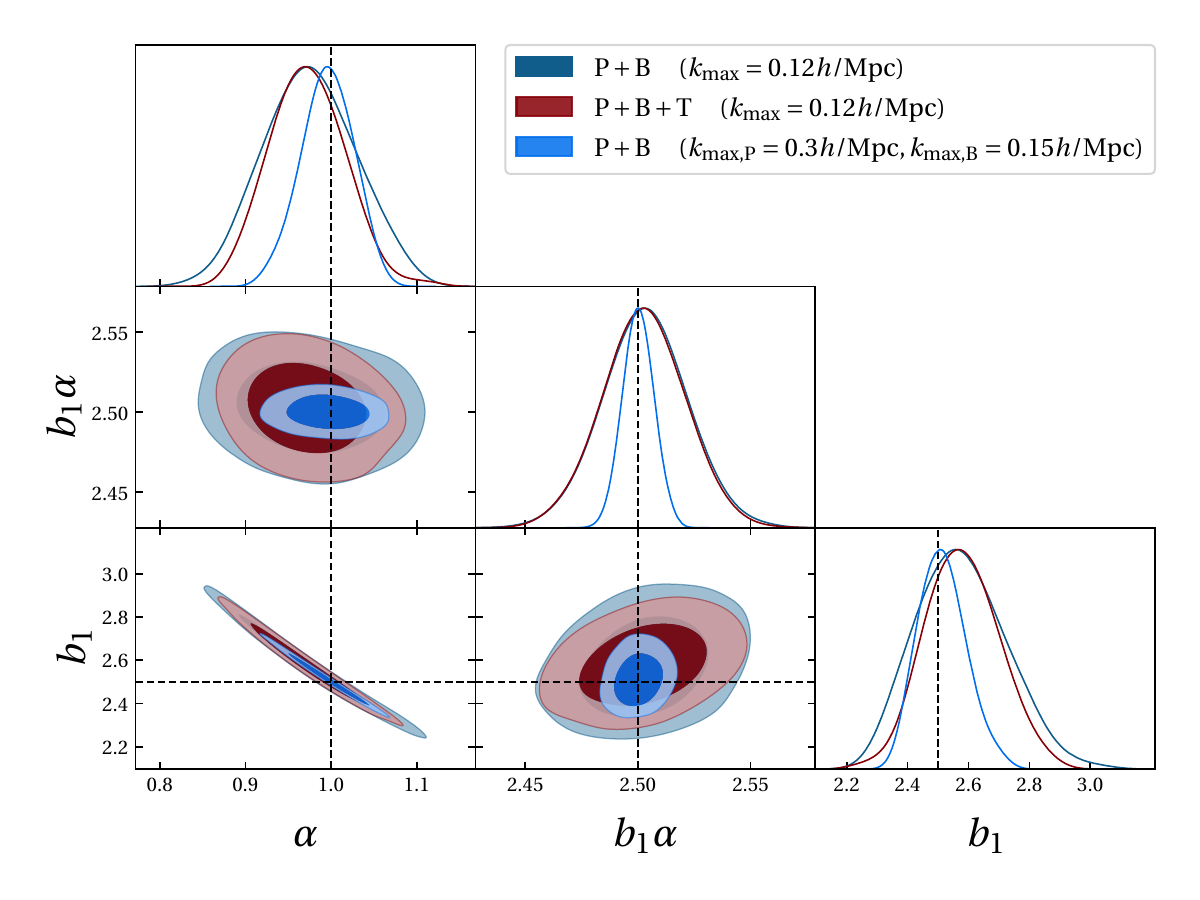}
    \caption{P+B and P+B+T analyses of the sample of halos similar to \cite{Beyond-2pt:2024mqz}. P+B results for $k_{\mathrm{max,P}}=0.3h/\mathrm{Mpc}, k_{\mathrm{max,B}}=0.15h/\mathrm{Mpc}$ fixing $b_{\Gamma_3},c_2,c_3$ to their fiducial values are also shown.}
    \label{fig:beyondPT}
\end{figure}
\begin{table}
    \centering
    \begin{tabular}{|c|c|c|c|} \hline 
         &   P+B&  P+B+T& P+B ($k_{\mathrm{max,P}}=0.3h/\mathrm{Mpc}, k_{\mathrm{max,B}}=0.15h/\mathrm{Mpc}$)\\ \hline 
         $\alpha$&   0.975$\pm$0.055&  0.973$\pm$0.045& 0.995$\pm$0.031\\ \hline 
         $b_1$&   2.575$\pm$0.151&  2.578$\pm$0.125& 2.514$\pm$0.0810\\ \hline
    \end{tabular}
    \caption{$1\sigma$ constraints and best fit parameters for P+B and P+B+T with~$k_{\mathrm{max}}=0.12\,h/\mathrm{Mpc}$ (first and second columns) and P+B with~$k_{\mathrm{max,P}}=0.3\,h/\mathrm{Mpc}$ and~$k_{\mathrm{max,B}}=0.15\,h/\mathrm{Mpc}$ (third column).}
    \label{tab:beyondPT}
\end{table}

\subsection{Comparison to ``Beyond 2pt challenge'' results}
\label{app:beyond2pfchallenge}

\begin{figure}
    \centering
    \includegraphics[width=0.65\linewidth]{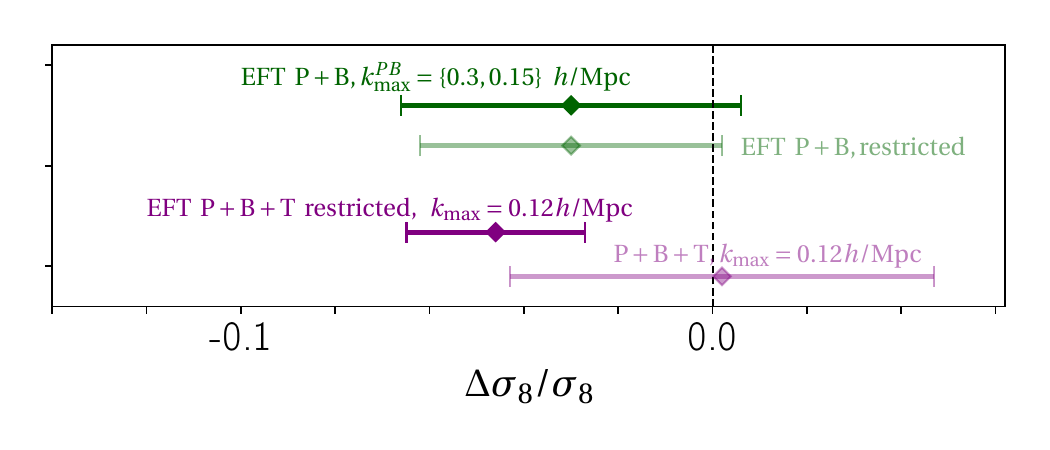}
    \caption{Error bars on the amplitude of the liner density field obtained in different analyses described in the main text. This figure is supposed to be compared with Fig.~22 in~\cite{Beyond-2pt:2024mqz}. The central values are chosen to match the Fig.~22 in~\cite{Beyond-2pt:2024mqz}. While our error bars are  different at~$\mathcal O(20\%)$ level, the general trend for all example is the same.}
    \label{fig:beyondPT_err}
\end{figure}

\noindent We now apply the same methodology to a sample which resembles the one analyzed in~\cite{Beyond-2pt:2024mqz}. Note that the simulation volume is the same~$V=(2\,\mathrm{Gpc}/h)^3$ but at slightly higher redshift~$z=1$. Using the exact same pipeline, we perform four different analyses. The results are shown in~Fig.\ref{fig:beyondPT},~Fig.\ref{fig:beyondPT_err} and Tab.~\ref{tab:beyondPT}.
\begin{itemize}
    \item First, we perform a joint P+B analysis with~$k_{\mathrm{max,P}}=0.3\,h/\mathrm{Mpc}$ and~$k_{\mathrm{max,B}}=0.15\,h/\mathrm{Mpc}$. We do two versions of this analysis. In one of them we fix~$b_{\Gamma_3},c_2,c_3$ to their fiducial values. This setup is exactly the same as the EFT P+B ``restricted'' analysis in~\cite{Beyond-2pt:2024mqz} (see Fig.~22), with the same choice of prior. We recover very similar error bar on~$\alpha$ of~$3\%$. In the other version we also vary~$b_{\Gamma_3},c_2,c_3$ within their usual priors. This corresponds to the baseline EFT P+B analysis in~\cite{Beyond-2pt:2024mqz}. In this case we find a~$\sim4\%$ error, which is smaller than what is reported in~\cite{Beyond-2pt:2024mqz} . However, given that our results are uncertain at the~$\mathcal O(20\%)$ level, we achieve a relatively good agreement with the EFT P+B analyses in~\cite{Beyond-2pt:2024mqz}. This serves as a cross check that our pipeline works well and that our choice of fiducial parameters is reasonable.
    \item Second, we perform the joint P+B analysis with~$k_{\mathrm{max}}=0.12\,h/\mathrm{Mpc}$ using priors given in Eq.~\eqref{eq:priorsforsample1}. This scale cut corresponds to~$k_{\mathrm{max}}=0.1\,h/\mathrm{Mpc}$ of~\cite{Beyond-2pt:2024mqz} where the cubic filter in~$k$-space is used. Since we use the spherical momentum shells, the two choices have a comparable number of Fourier modes (for more details see~\cite{Beyond-2pt:2024mqz}). The error on~$\alpha$ in this case is~$5.5\%$, approximately~$30\%$ larger than in the FBI ``extended'' analysis of~\cite{Beyond-2pt:2024mqz}. This is similar to the results we found in other cases and indicates the broad agreement between the FBI and P+B analyses. 
    \item Third, we add the trispectrum and perform the joint P+B+T analysis with the same cut~$k_{\mathrm{max}}=0.12h/\mathrm{Mpc}$. Similarly to what happens for other tracers, also here we find an improvement of roughly~$20\%$ compared to the P+B analysis. The joint P+B+T result is in rough agreement with the FBI error of~$4\%$ obtained in~\cite{Beyond-2pt:2024mqz}.
     \item Finally,  we run a P+B+T analysis at~$k_{\mathrm{max}}=0.12\,h/\mathrm{Mpc}$ but this time fixing all cubic biases and the bispectrum noise terms. In other words, the only free parameters remaining are the linear and quadratic biases, the dark matter counterterm and the power spectrum shot noise. We choose this setup to be as close as possible to the FBI ``restricted'' analysis in~\cite{Beyond-2pt:2024mqz}. In this case we find a significantly smaller error on~$\alpha$ of~$2\%$. This matches nicely the FBI ``restricted'' analysis in~\cite{Beyond-2pt:2024mqz}.
     \end{itemize}

The last example is particularly interesting. If we compare the ``restricted'' P+B+T and P+B  analyses at the same~$k_{\rm max}=0.12\;h/{\rm Mpc}$, we find that the addition of the trispectrum when cubic biases are fixed makes the constraints better by more than a factor of 2. Note that the impact of fixing cubic biases in the P+B analysis is minimal. This is a very significant improvement. Does this mean that our intuition for the relevance of the trispectrum is wrong? We think not. It is still true that the trispectrum has much lower SNR than the power spectrum and the bispectrum. However, since the trispectrum in this particular setup comes with no new free parameters by design, even this small signal is very useful to break the residual degeneracy between~$\alpha$ and~$b_2$ and~$b_{\mathcal G_2}$ in the bispectrum. This degeneracy breaking is what leads to much smaller error bar on~$\alpha$. Clearly, this example is somewhat artificial and designed to maximize the impact of~$n$-point functions beyond the bispectrum. For a real tracer where the cubic and higher order biases are not zero, such approach is inappropriate, as evident from the biased result of the FBI ``restricted'' analysis in~\cite{Beyond-2pt:2024mqz}. Pushing this to the extreme, one can always design a setup or find a cosmological parameter such that the relevant information is only in the higher order correlation functions. For example, if we wanted to measure the primordial non-Gaussianity parameter~$\tau_{\rm NL}$, the FBI would be much much better than the P+B analysis. However, these special situations are not what we have in mind when we discuss the comparison of the field-level inference and the standard analyses. In a generic setup, our arguments are still valid. 
     
In order to summarize our results and visually compare them to analyses in~\cite{Beyond-2pt:2024mqz}, we produce Fig.~\ref{fig:beyondPT_err} which is supposed to be compared to the Fig.~22 in~\cite{Beyond-2pt:2024mqz}. We choose the central values of the intervals to match the results in~\cite{Beyond-2pt:2024mqz} and we plot our forecast for the error bars. While all our results are uncertain at the~$\mathcal O(20\%)$ level, we find an overall good agreement with results of~\cite{Beyond-2pt:2024mqz} and we recover all features and trends. These results are reassuring and provide additional evidence for the simple relation between the~$n$-point functions and the field level inference on large scales.  

In conclusion, all our results are in agreement with expectations from perturbation theory. For dark matter halos in real space, where the only cosmological parameter is the amplitude of the linear field~$\alpha$, most of the information comes from the bispectrum, except in some artificial examples where the higher order correlation functions can dominate the SNR. Even though the power spectrum has the highest signal-to-noise, due to the exact degeneracy between~$b_1$ and~$\alpha$ in linear theory, the amplitude can be measured from one-loop power spectrum only with a modest precision of roughly~$20\%$ in a box of volume~$V=8\,({\rm Gpc}/h)^3$ and with~$k_{\rm max}=0.12\; h/{\rm Mpc}$. On the other hand, adding the bispectrum improves these errors to approximately~$5-6\%$ (the exact number slightly depends on the sample). Addition of the trispectrum further reduces the error to roughly~$4-5\%$. These errors are in good agreement with the measurement of~$\alpha$ using FBI~\cite{Nguyen:2024yth,Beyond-2pt:2024mqz}.

For simplicity, in this section we focused on only one of the two samples studied in~\cite{Nguyen:2024yth}. However, we also checked that for the other sample at~$z=1$ and with even larger number density, we find very similar results as the FBI analysis and all our conclusions remain the same in that case as well.

\section{Discussion}
\label{sec:discussion}
\noindent The results presented in the previous section indicate that the conventional analyses based on correlation functions and the FBI analysis lead to the same results in the perturbative regime, as expected based on theoretical arguments~\cite{Cabass:2023nyo,Schmidt:2025iwa}. This seems to be in contradiction with the findings in~\cite{Nguyen:2024yth} where the FBI leads to up to 5 times better error bars than the SBI P+B analysis. The new puzzle is why the SBI and the standard P+B analyses differ so much. In this section we try to provide some possible explanations, even though the conclusive arguments are still lacking and more investigation is needed to fully clarify this issue. 

The key to understanding the different results is to understand the differences in the settings of the SBI analysis of~\cite{Nguyen:2024yth} and the settings in this paper. There are three major differences. One is the theoretical model used. SBI analysis of~\cite{Nguyen:2024yth} is based on LPT, while we use the Eulerian theory. As we argued in detail in this work and as has been checked in the literature before, the two models are practically indistinguishable on the relevant scales and we do not think that this can be the source of the disagreement. The second major difference is the likelihood used in the analyses. SBI does not assume any likelihood (even though it has been shown that a Gaussian likelihood is an excellent approximation~\cite{Tucci:2023bag}) and uses a large number of realizations to estimate the relevant scatter in the bins for the data vector. On the other hand, we use a simple Gaussian likelihood with the Gaussian covariance, neglecting the cross-covariance between P, B and T. We have also justified this choice in the paper and there is numerous evidence in the literature that this is a good approximation. Finally, the third major difference is the choice of bias parameters in the analyses and the treatment of priors on these nuisance parameters. We believe that this can be the main source of disagreement. 

In this work we follow the usual logic used in perturbation theory and analyses based on the~$n$-point functions. First, given the statistical error bars of the data, we estimate to what perturbative order we have to compute observables in order to be unbiased. In our examples, such estimate indicates that the one-loop power spectrum and the tree-level bispectrum are sufficient. Once the order in PT is fixed, the bias parameters that have to be used in the analysis are also determined. It is important to emphasize that here we do not truncate the bias expansion, we are only truncating the loop expansion of the correlation functions, given the total SNR that we expect for a given data set. For example, in the case where we compute power spectrum up to one-loop and the bispectrum at tree level, even though we use the full cubic bias model, only one of the cubic biases appears explicitly in the one-loop power spectrum~($b_{\Gamma_3}$). The other cubic biases either give zero contribution at this order in PT or just renormalize~$b_1$. The complexity of the nonlinear field at this order in PT and for the simplest observables such as P and B is just not sufficient for other bias parameters to appear explicitly. Obviously, cubic biases would appear if we had gone to higher order in the loop expansion or higher~$n$-point functions. However, since we estimate that they are not necessary to fit the data, we can safely drop these refinements of the basic perturbative prediction. To conclude, even though our model is based on the full bias expansion up to the cubic order, 3 out of 4 cubic bias parameters do not appear in the analysis. This is a consequence of bias renormalization and the loop expansion and not restricting the number of bias parameters that are used in the analysis. 

On the other hand, the practical implementation of SBI in~\cite{Nguyen:2024yth} uses the nonlinear field computed at cubic order with all relevant bias parameters to predict P and B. This is similar in spirit to the field-level inference, where one can only truncate the forward model to some order in the bias expansion, without the possibility to explicitly restrict the likelihood to certain correlation functions or number of loops. Therefore, the estimated power spectrum and bispectrum contain higher order contributions beyond what we use in our model. For example, the power spectrum has a part of the 2-loop terms, such as various~$P_{33}$ contributions. The bispectrum has a part of the one-loop contributions (diagrams usually labeled as~$B_{123}$ and~$B_{222}$) and even some two-loop contributions such as~($B_{233}$). It is interesting to explore whether this difference leads to a large difference in the error bars in the two P+B analyses. 

Naively, one may think that this difference is irrelevant, since the higher order terms are small in PT. However, this is not obvious. While the variance of the dark matter field is small for~$k_{\rm max}$ used in the analysis, the fact that there are many contributions from the bias expansion means that there are combinations of bias parameters that might produce a large effect. With all bias parameters uncorrelated and of order one, such combinations are naturally explored in the MCMC chains. Marginalizing over such parameters can then make the error bars on~$\alpha$ significantly higher than the standard expectation. We do not have a clear evidence that this is the main origin of the disagreement, but it is certainly an effect to be carefully taken into account in any comparison. 

In order to illustrate this point better, let us imagine that we keep going to higher and higher orders in PT hoping to improve the precision of our inference. If we keep~$k_{\rm max}$ fixed, this means that we fit the same data with a theory that has a rapidly increasing number of free parameters. Without careful choice of the priors, even though the individual loop diagrams may be small, summing a large number of such terms would inevitably lead to large effects. In such case the {\em total} loop contribution stops being small and the perturbativity in this sense is lost. The details of when this happens for a given~$k_{\rm max}$ and the given observable depend on the sample and the priors, but one always has to check if this is the case. The hints of this behavior were already noticed in the analyses of the one-loop bispectrum~\cite{Philcox:2022frc,DAmico:2022osl,Braganca:2023pcp}. On the other hand, for the consistency of PT, adding higher order terms should at worst leave the constraints unchanged, it should never make them worse.  

What are the usual ways to mitigate this problem? One possible procedure is to add informative and physically motivated priors on bias parameters which makes them correlated. These correlations can be estimated from some external analysis (see for example~\cite{Akitsu:2024lyt,Ivanov:2024hgq,Ivanov:2024xgb,Zhang:2024thl,Ivanov:2025qie} for HOD informed priors obtained from a simulation-based approach and from analytical predictions respectively).
Another possible solution is to use the so-called perturbativity priors~\cite{Braganca:2023pcp}, which we already mentioned in the introduction. These theoretically motivated priors do not rely on any galaxy evolution physics and are imposed in order to control the size of the loop corrections predicted by PT (see~\cite{Braganca:2023pcp} for more details). It is important to emphasize that these two procedures lead to similar results, indicating that the real tracers indeed never produce large fluctuations due to bias nonlinearities. Perturbativity priors are also one implicit assumption that we always make for perturbation theory to make sense. Note that requiring that fluctuations in the galaxy density field are small on large scales still allows that all biases are of order 1, but imposes correlations among them, similar to the ones found in HOD or hydrodynamical simulations. 

If one accepts the perturbativity prior, what are the consequences for the SBI analysis? In principle, one would have to restrict the sampling only to those values of biases that do not produce large loop corrections. In this case, the expectation is that the cubic terms in the bias expansion would be much less relevant and the SBI will be similar to the standard analysis of the tree-level bispectrum presented in this paper. This intuition provides a simple way to test the relevance of cubic terms. For example, one can set all cubic parameters to zero in the SBI analysis and try to recover the standard results. The other easy check is to plot the one-loop bispectrum amplitudes as a function of scale for various shapes (similar to Fig.~\ref{fig:bisp_fit}) for all samples in the chains in the SBI and see if the envelope they create coincides with the estimate of the theoretical error.  

If we assume that the wide priors are the origin of the disagreement between the SBI and the standard analysis, one may wonder why the FBI still gives good results with presumably large fluctuations in the galaxy density field. The answer is that the FBI is always optimal if the model is close enough to the data. For the combination of biases that do not respect perturbativity the model would contain a large fluctuation in the galaxy density field with the particular template given by PT. FBI compares such model with the whole field and assigns a small likelihood to such set of bias parameters. Therefore, there is no contradiction in having small errors in the FBI approach and large errors from P+B if the nuisance parameters are degenerate enough in the P+B combination. 

In conclusion, one possible origin of disagreement can be that including higher-order loop contributions on very large scales where they are supposed to be negligible can lead to artificial increase in the variance of the inferred cosmological parameters. In order to avoid this, one can either do the lower order calculation as we did in this paper, or impose some perturbativity priors in the inference. If one does either of these, the results obtained using~$n$-point functions at fixed loop order and the SBI analysis should lead to similar constraints. Luckily, this can be relatively easily tested and we leave this for future work.

\section{Conclusions}
\label{sec:conclusions}

\noindent In this paper we studied a simple setup of dark matter halos in real space on very large scales where the only free cosmological parameter is the amplitude of the linear density field~$\alpha$. We computed the errors on~$\alpha$ in the joint analyses of P+B and P+B+T. We confirmed that most of the information comes from the bispectrum which breaks the~$b_1-\alpha$ degeneracy and that the addition of the trispectrum improves the error only by~$20-30\%$. Our results for the P+B joint analysis are in agreement with previous literature. This indicates that the information on large scales is saturated by a few leading order~$n$-point functions and dominated by the bispectrum. 

These results are important in the context of comparison to alternative data analysis techniques, such as the field-level inference of cosmological parameters. Our work suggests that the error bars in the P+B+T analysis are in good agreement with the recent FBI results. In the absence of some accidents and coincidences, this implies the clear connection between the FBI and the conventional analyses and confirms the theoretical expectation that the field-level analysis in perturbative forward modeling is ultimately equivalent to measuring the~$n$-point correlation functions computed at the same perturbative order. Of course, FBI is by construction optimal and, assuming everything else the same, it will always lead to slightly better errors than the correlation function based analyses. However, this improvement for perturbative forward modeling should be small and certainly not as significant as a factor of 2 or more. 

One open problem that remains to be solved is the difference of the PT based P+B analysis presented in this paper and the SBI P+B analysis of~\cite{Nguyen:2024yth}. While we do not have any definitive conclusions, we gave some possible explanations for the origin of the discrepancy in the inferred error bars. It would be very important to resolve this problem since it touches upon very important questions of how to apply PT in practice and how to choose appropriate priors for nuisance parameters. All these questions will be progressively more important as we move on towards more complicated observables and higher order computations in PT. 

While throughout most of this paper we focused on the typical galaxy sample where the FBI is expected to be equivalent to the standard correlation functions based analysis, it is important to emphasize that there are situations where the FBI is supposed to perform better. This includes the BAO reconstruction~\cite{Babic:2022dws,Babic:2024wph,Babic:2025fgv} or a very high number density tracer, such as neutral hydrogen~\cite{Obuljen:2022cjo,Foreman:2024kzw,Kokron:2025yma}. In such examples there are new large parameters, beyond the variance of the density field, that suppress the signal in the conventional analysis either by erasing the information (BAO) or making the covariance large (dense tracers). It would be interesting to repeat our analysis for these particular setups and show the difference compared to FBI more explicitly. 

Apart from the comparison with the field-level inference, our work is also important since it provides a simple analytical model for the trispectrum of biased tracers in real space and its covariance matrix. This extends previous works that have used some compressed observables, such as the integrated trispectrum~\cite{Gualdi:2020eag,Gualdi:2022kwz}. Our formalism can be used to get a quick estimates of the error bars for the signals specific to the 4-point function, such as models that produce parity violation (for example~\cite{Cabass:2022oap}) or non-conventional shapes of primordial non-Gaussianities (for example~\cite{Creminelli:2012qr}). Furthermore, our likelihood can be used to explore if there is a compression of the trispectrum that preserves most of the information while significantly reducing the dimensionality of the data vector~\cite{Heavens:1999am,Scoccimarro:2000sn,Reichardt:2001ku,Eisenstein:2006nk,Philcox:2020zyp}. Such compression would significantly simplify realistic data analyses. 

At the end, let us point out that all our results are derived for the real space dark matter halos and a single cosmological parameter~$\alpha$. While this is an important step to conceptually understand the relation between the field-level inference and conventional analyses based on the~$n$-point functions, it is unclear to what extent the lessons we learn transfer to the more realistic case of redshift space observables and other cosmological parameters. For instance, the~$b_1-\alpha$ degeneracy in redshift space is partially broken by the power spectrum quadrupole. For this reason, in the absence of strong observational bias effects~\cite{Desjacques:2018pfv},  the results would be different than for the real space. For example, it is found in practice that the addition of the bispectrum in the analysis of the redshift space data does not significantly improve the error on~$\alpha$. Therefore, the contributions to the SNR from different~$n$-point functions are different in redshift space and it would be interesting to repeat our analysis there. Similarly, other cosmological parameters have different degeneracies with biases and it would be interesting to explore to what extent our general arguments apply in that case as well.

\section*{Acknowledgments}
\noindent We thank Kazuyuki Akitsu, Giovanni Cabass, Michele Liguori, Massimo Pietroni, Alvise Raccanelli and Beatriz Tucci for useful discussions. We also thank Nhat-Minh Nguyen, Fabian Schmidt and Beatriz Tucci for many useful comments on the manuscript. We thank Takahiro Nishimichi for providing the data of the PT Challenge simulations.
MM acknowledges support from MUR Progetti di Ricerca di Rilevante Interesse Nazionale (PRIN) Bando 2022 - grant 20228RMX4A, funded by the European Union - Next generation EU, Mission 4, Component 1, CUP C53D23000940006.

\appendix

\section{Cross-correlation between EPT and LPT}
\label{app:EPTvsLPT}

\noindent The cross-correlation between the leading order EPT and LPT fields is given by
\begin{align}\label{eq:deriv1}
\langle \delta_{\rm EPT}^{\rm LO}(\k_1) \delta_{\rm LPT}^{\rm LO}(\k_2) \rangle & = \int d^3 \q_1 d^3 \q_2 \, \langle \delta_1(\q_1) e^{-i\k_2\cdot\boldsymbol\psi_1(\q_2)} \rangle \, e^{-i\k_1\cdot\q_1- i\k_2\cdot\q_2} \nonumber \\
& = i \frac{d}{d\lambda} \int d^3 \q_1d^3 \q_2 \, \langle e^{-i(\lambda \delta_1(\q_1) +\k_2\cdot\boldsymbol\psi_1(\q_2))} \rangle \, e^{-i\k_1\cdot\q_1 - i\k_2\cdot\q_2} \Big|_{\lambda=0} \;.
\end{align}
The expectation value can be computed using the cumulant theorem. For a random variable $X$, we have
\begin{equation}
\langle e^{-iX} \rangle = \exp \left( \sum_{n=0}^\infty \frac{(-i)^n}{n!} \langle X^n \rangle_c \right) \;,
\end{equation}
where the cumulants $\langle X^n \rangle_c$ in our case represent the connected $n$-point functions. Since at LO all fields in the exponent are Gaussian, the only nontrivial cumulant is the power spectrum. After taking the derivative and setting $\lambda=0$ we get
\begin{align}
i \frac{d}{d\lambda} \langle e^{-i(\lambda \delta_1(\q_1) +\k_2\cdot\boldsymbol\psi_1(\q_2))} \rangle \Big|_{\lambda=0} & = i \frac{d}{d\lambda} e^{-\frac 12 \left( \lambda^2 \langle \delta_1^2(\q_1) \rangle + 2 \lambda \k_2 \cdot \langle \delta_1(\q_1) \boldsymbol\psi_1(\q_2) \rangle + k_2^ik_2^j \langle \psi_{1i}(\q_2) \psi_{1j}(\q_2) \rangle \right) } \Big|_{\lambda=0} \nonumber \\
& = -i \k_2 \cdot  \langle\delta_1(\q_1)\boldsymbol \psi_1(\q_2)\rangle e^{-\frac{1}{2}k_2^2\sigma_v^2 } \;.
\end{align}
Using this in the Eq.~\eqref{eq:deriv1} gives the following exact result for the LO cross correlation 
\begin{equation}
\langle \delta_{\rm EPT}^{\rm LO}(\k) \delta_{\rm LPT}^{\rm LO}(-\k) \rangle' = P_{\rm lin}(k) e^{-\frac{1}{2}k^2\sigma_v^2} \;.
\end{equation}
Assuming that at LO all power spectra are approximately equal to the linear power spectrum on large scales, we get the following LO cross-correlation coefficient
\begin{equation}
r^{\rm LO}(k)  \approx e^{-\frac{1}{2}k^2 \sigma_v^2} \;,
\end{equation}
which is the Eq.~\eqref{eq:cross_linear} in the main text.

The NLO cross-correlation can be derived in a similar way. It is clear from the generic structure of the LPT expressions that it must be exponentially suppressed on small scale, but keeping the cubic terms in the Eulerian perturbation theory guaranties partial cancellation of large displacement effects on large scales.  

\section{Trispectrum shot noise terms}
\label{app:noise}
\noindent Here we show all the possible contributions to the trispectrum shot noise. The total galaxy field can indeed be expressed as a sum of the perturbative solution and the stochastic contribution
\begin{equation}
    \delta_g(\x) = \delta_g^{\rm PT}(\x) + \epsilon(\x)\;.
    \label{eq:App_dtot}
\end{equation}
We expand the noise field using 
\begin{equation}
\epsilon(\x)= \epsilon_0(\x) + \epsilon_1(\x) \delta(\x) + \frac 12 \epsilon_2(\x)(\delta^2(\x)-\langle \delta^2 \rangle) + \epsilon_{\mathcal G_2}(\x) \mathcal G_2 (\x) + \ldots \;,
\end{equation}
and, since it is well-known that the spectra of the noise in the Poissonian limit satisfy
\begin{equation}
    B_{\epsilon} = P^2_\epsilon = \frac{1}{\bar{n}_g^2}\;,
\end{equation}
we define the correlation among different noise operators as
\begin{align}
\langle\epsilon_i(\x)\epsilon_j(\x')\rangle &\equiv \frac{c_{ij}}{\bar{n}_g}\delta^D(\x - \x')\,, \label{eq:App_cij}\\
\langle\epsilon_i(\x_1)\epsilon_j(\x_2)\epsilon_k(\x_3)\rangle &\equiv \frac{c_{ijk}}{\bar{n}_g^2}\delta^D(\x_1 - \x_2)\delta^D(\x_2 - \x_3)\,, \label{eq:App_cijk}\\
\langle\epsilon_i(\x_1)\epsilon_j(\x_2)\epsilon_k(\x_3)\epsilon_l(\x_4)\rangle &\equiv \frac{c_{ijkl}}{\bar{n}_g^3}\delta^D(\x_1 - \x_2)\delta^D(\x_2 - \x_3)\delta^D(\x_3 - \x_4)\;. \label{eq:App_cijkl}
\end{align}
As a warm-up, we compute the power spectrum and bispectrum stochastic terms: the power spectrum of Eq.~\eqref{eq:App_dtot} is
\begin{equation}
    P_{gg}(\k) = \langle\delta_g(\k)\delta_g(\k')\rangle' + \langle\epsilon(\k)\epsilon(\k')\rangle'\;,
\end{equation}
where we have exploited $\langle\delta \epsilon\rangle = 0$. The leading stochastic contribution for the galaxy power spectrum is then
\begin{equation}
    \langle\epsilon_0 (\k)\epsilon_0 (\k')\rangle' = \frac{c_{00}}{\bar{n}_g}\;,
\end{equation}
with $c_{00}$ being defined in Equation~\eqref{eq:App_cij}. Notice that additional terms in the expansion could arise that will eventually renormalize the $c_{00}$ coefficient, e.g. the term $\langle \epsilon_1(\x)\delta(\x)\epsilon_1(\x')\delta(\x')\rangle'$. For the bispectrum the leading shot noise terms are
\begin{equation}
    \langle \delta_g^{\rm PT}(\k_1)\epsilon(\k_2)\epsilon(\k_3)\rangle' + 2 \;\text{perms.} = \frac{c_{01}}{\bar{n}_g} b_1\big(P(k_1) +P(k_2) + P(k_3)\big)\;,
\end{equation}
and 
\begin{equation}
\langle\epsilon(\k_1)\epsilon(\k_2)\epsilon(\k_3)\rangle' = \frac{c_{000}}{\bar{n}_g^2}\;.
\end{equation}
For the trispectrum we can write all the possible terms in the perturbative expansion as
\begin{align}
\label{eq:App_Ttot}
T_g(\k_1,\k_2,\k_3,\k_4)=& \langle\delta_g(\k_1)\delta_g(\k_2)\delta_g(\k_3)\delta_g(\k_4)\rangle' \\
=& \langle\delta_g^{\rm PT}(\k_1)\delta_g^{\rm PT}(\k_2)\delta_g^{\rm PT}(\k_3)\delta_g^{\rm PT}(\k_4)\rangle' \nonumber\\
&+ \langle\delta_g^{\rm PT}(\k_1)\delta_g^{\rm PT}(\k_2)\delta_g^{\rm PT}(\k_3)\epsilon(\k_4)\rangle' + 3 \; \text{perms}\nonumber\\
& + \langle\delta_g^{\rm PT}(\k_1)\delta_g^{\rm PT}(\k_2)\epsilon(\k_3)\epsilon(\k_4)\rangle' + 5 \; \text{perms}\nonumber\\
& + \langle\delta_g^{\rm PT}(\k_1)\epsilon(\k_2)\epsilon(\k_3)\epsilon(\k_4)\rangle' + 3 \; \text{perms}\nonumber\\
& + \langle\epsilon(\k_1)\epsilon(\k_2)\epsilon(\k_3)\epsilon(\k_4)\rangle' \;.\nonumber
\end{align}
The first line of Equation~\eqref{eq:App_Ttot} is simply the tree-level perturbative expression for the galaxy trispectrum. The second line is automatically zero, since $\langle\epsilon\rangle = 0$. The third line gives rise to a number of possible stochastic terms (up to 1-loop order):
\begin{align}
    &\langle\delta_g^{\rm PT}(\k_1)\delta_g^{\rm PT}(\k_2)\epsilon(\k_3)\epsilon(\k_4)\rangle' = \langle b_1 \delta(\k_1)b_1 \delta(\k_2)\epsilon_0(\k_3)\epsilon_0 (\k_4)\rangle'\nonumber\\
    &+ \langle b_1\delta(\k_1)b_1\delta(\k_2)\epsilon_0(\k_3)\int_\p\epsilon_1(\p)\delta(\k_4 - \p)\rangle' +  \langle b_1\delta(\k_1)b_1\delta(\k_2)\epsilon_0(\k_3)\int_\p\epsilon_2(\p)\delta^2(\k_4 - \p)\rangle' \nonumber\\
    &+  \langle b_1\delta(\k_1)b_1\delta(\k_2)\epsilon_0(\k_3)\int_\p\epsilon_{\mathcal{G}_2}(\p)\mathcal{G}_2(\k_4 - \p)\rangle' +  \langle b_1\delta(\k_1)b_1\delta(\k_2)\int_\p\epsilon_1(\p)\delta(\k_3 - \p)\int_\p\epsilon_1(\p)\delta(\k_4 - \p)\rangle'\\
    & \langle b_1 \delta(\k_1)b_2\delta^2(\k_2)\epsilon_0(\k_3)\epsilon_0(\k_4)\rangle' + \langle b_1 \delta(\k_1)b_{\mathcal{G}_2}\mathcal{G}_2(\k_2)\epsilon_0(\k_3)\epsilon_0(\k_4)\rangle'\nonumber\\
    & + \langle b_1 \delta(\k_1)b_2 \delta^2(\k_2)\epsilon_0(\k_3)\int_\p \epsilon_1(\p)\delta(\k_3 - \p)\rangle' + \langle b_1 \delta(\k_1)b_{\mathcal{G}_2}\mathcal{G}_2(\k_2)\epsilon_0(\k_3)\int_\p \epsilon_1(\p)\delta(\k_3 - \p)\rangle'\;,\nonumber
\end{align}
that eventually give the $\propto \bar{n}_g^{-1}$ contribution
\begin{equation}
    \langle\delta_g^{\rm PT}(\k_1)\delta_g^{\rm PT}(\k_2)\epsilon(\k_3)\epsilon(\k_4)\rangle' = \frac{c_{01}}{\bar{n}_g}B_{mgg}(\k_{12},\k_3, \k_4) + \frac{2 b_1^2}{\bar{n}_g}\left(c_{02} + c_{0\mathcal{G}_2}\sigma^2(\k_1, \k_2) + c_{11}\right) P(k_1) P(k_2)\;.
\end{equation}
The fourth line of Equation~\eqref{eq:App_Ttot} produces contributions of order $\sim \bar{n}_g^2$
\begin{align}
    \langle\delta_g^{\rm PT}(\k_1)\epsilon(\k_2)\epsilon(\k_3)\epsilon(\k_4)\rangle' \simeq \frac{b_1 c_{001}}{\bar{n}_g^2}P(k_1) + 3 \, \text{perms.}\;. 
\end{align}
at leading perturbative order.
The last contribution is given by the noise trispectrum
\begin{equation}
\langle\epsilon(\k_1)\epsilon(\k_2)\epsilon(\k_3)\epsilon(\k_4)\rangle' = \frac{c_{0000}}{\bar{n}_g^3}\;.
\end{equation}

\section{Estimators and covariance matrices}
\label{app:covariances}

\noindent In this appendix we derive the expression for the trispectrum covariance matrix. The trispectrum configurations are labeled by quadrilaterals $Q$ defined by the four momenta and the two diagonals $Q=\{ k_1, k_2, k_3, k_4, D_1, D_2 \}$. In order to avoid  double counting and take into account all possible quadrilaterals, we have to order the momenta $k_1\geq k_2 \geq k_3 \geq k_4$ but we leave the diagonals free, subject to the constraint that the first diagonal is always the sum of the two largest momenta $\boldsymbol{D_1}+\k_1+\k_2=0$ and the second diagonal is always the sum of second and third largest momenta $\boldsymbol{D_2}+\k_2+\k_3=0$. The estimator for the connected trispectrum can be then written as 
\begin{align}
\hat T_c(k_1,k_2,k_3,k_4,D_1,D_2) & = \frac1{VV_{1234}} \int_{\q_i\in k_i} \int_{\p_i\in D_i} \hat\delta_g(\q_1) \hat\delta_g(\q_2) \hat\delta_g(\q_3)
\hat\delta_g(\q_4) \,\delta^D(\q_{1234}) \delta^D(\q_{12}+\p_1) \delta^D(\q_{23}+\p_2) \nonumber \\
& \qquad - {\rm disconnected \; terms} \;,
\end{align}
where the ``disconnected terms'' are chosen such that the estimator is unbiased $\langle \hat T_c \rangle = T_c$. Their explicit form is not of interest in the Gaussian likelihood, since they cancel in the covariance matrix.

Let us first compute the volume for the thin momentum shell containing all quadrilaterals with given momenta and diagonals. It is defined as
\beq
V_{1234} = \int_{p_i\in D_i} d^3 \p_i \int_{q_j \in k_j} d^3 \q_j \, \delta^D(\q_1+\q_2+\q_3+\q_4) \delta^D(\q_1+\q_2+\p_1) \delta^D(\q_2+\q_3+\p_2) \; .
\eeq
Note that the symmetry among six edges of the tetrahedron is broken the moment these particular delta functions are chosen. What complicates the evaluation of the integral is that~$\q_2$ appears three times as an argument of delta functions. When doing angular integrals this will lead to Wigner 3j symbols. Therefore, it is more convenient to rewrite the previous formula as follows
\beq
V_{1234} = \int_{p_i\in D_i} d^3 \p_i \int_{q_j \in k_j} d^3 \q_j \,  \delta^D(\q_1+\q_4-\p_2) \delta^D(\q_1+\q_2+\p_1) \delta^D(\q_2+\q_3+\p_2) \;,
\eeq
which we can always do using the delta functions. This form is simpler, since any of the momenta enter the arguments of the delta functions at most twice. Next we can write
\beq
V_{1234} = \int_{p_i\in D_i} d^3 \p_i \int_{q_j \in k_j} d^3 \q_j \int \frac{d^3\r}{(2\pi)^3} \frac{d^3\x}{(2\pi)^3} \frac{d^3\y}{(2\pi)^3} \, e^{- i\r \cdot(\q_1+\q_4-\p_2) -i\x \cdot (\q_1+\q_2+\p_1) - i\y \cdot(\q_2+\q_3+\p_2) } \;.
\eeq 
Angular integration over $\p_1$, $\q_4$ and $\q_3$ can be done easily. 
\begin{align}
V_{1234} & = \frac{1}{8\pi^6} \int_{p_i\in D_i} p_i^2 dp_i  \int_{q_j \in k_j} q_j^2 dq_j  \int_0^\infty r^2 dr \int_0^\infty x^2 dx \int_0^\infty y^2 dy \, j_0(p_1x) j_0(q_4 r) j_0(q_3 y) \nonumber \\
& \qquad \times \int d\Omega_{p_2} \int d\Omega_{q_1} \int d\Omega_{q_2} \int d\Omega_{r} \int d\Omega_{x} \int d\Omega_{y}  \, e^{- i\r \cdot(\q_1-\p_2) -i\x \cdot (\q_1+\q_2) - i\y \cdot(\q_2+\p_2) } \; .
\end{align}
In order to perform the other angular integrals, we will use the expansion of plane waves in spherical Bessel functions and spherical harmonics
\beq
e^{\pm i \r \cdot \q} = 4\pi \sum_{\ell m} (\pm i)^\ell j_\ell (q r) Y_{\ell m}(\hat{\r}) Y^*_{\ell m}(\hat{\q}) \;. 
\label{eq:exp}
\eeq
It follows that
\beq
\int d\Omega_{q_1} e^{- i \r \cdot \q_1 - i\x \cdot \q_1} = (4\pi)^2 \sum_{\ell_1 m_1} \sum_{\ell_2 m_2} (- i)^{\ell_1+\ell_2} j_{\ell_1} (q_1 r) j_{\ell_2} (q_1 x) Y_{\ell_1 m_1}(\hat{\r}) Y_{\ell_2 m_2}^*(\hat{\x}) \int d\Omega_{q_1} Y^*_{\ell_1 m_1}(\hat{\q}_1) Y_{\ell_2 m_2}(\hat{\q}_1) \;. 
\eeq
By orthonormality of the spherical harmonic this integral leads to
\beq
\int d\Omega_{q_1} e^{- i \r \cdot \q_1 - i\x \cdot \q_1} = (4\pi)^2 \sum_{\ell m} (-1)^{\ell} j_{\ell} (q_1 r) j_{\ell} (q_1 x) Y_{\ell m}(\hat{\r}) Y_{\ell m}^*(\hat{\x}) \;. 
\eeq
We can to the other angular integrals in the similar way. In particular,
\beq
\int d\Omega_{q_2} e^{- i \x \cdot \q_2 - i\y \cdot \q_2} = (4\pi)^2 \sum_{\ell m} (-1)^{\ell} j_{\ell} (q_2 x) j_{\ell} (q_2 y) Y_{\ell m}(\hat{\x}) Y_{\ell m}^*(\hat{\y}) \;,
\eeq
\beq
\int d\Omega_{p_2} e^{+ i \r \cdot \p_2 - i\y \cdot \p_2} = (4\pi)^2 \sum_{\ell m} j_{\ell} (p_2 r) j_{\ell} (p_2 y) Y_{\ell m}(\hat{\y}) Y_{\ell m}^*(\hat{\r}) \;. 
\eeq
Once all the angular integrals are done, the volume of the shell is a complicated radial integral of product of spherical Bessel functions
\begin{align}
V_{1234}  & = 8^3 \int_{p_i\in D_i} p_i^2 dp_i  \int_{q_j \in k_j} q_j^2 dq_j  \int_0^\infty r^2 dr \int_0^\infty x^2 dx \int_0^\infty y^2 dy \sum_{\ell} (2\ell+1) \,  \nonumber \\
& \qquad \times j_0(p_1x) j_0(q_4 r) j_0(q_3 y) j_{\ell} (q_1 r) j_{\ell} (q_1 x) j_{\ell} (q_2 x) j_{\ell} (q_2 y) j_{\ell} (p_2 r) j_{\ell} (p_2 y) \; .
\end{align}
Luckily, the radial integrals can be also done relatively straightforwardly. First, we can use 
\beq\label{eq:legendre}
\int_0^\infty r^2 dr \, j_0(q_4 r) j_{\ell} (q_1 r) j_{\ell} (p_2 r) = \frac{\pi}{4q_1q_4p_2} \mathcal P_\ell \left( \frac{q_1^2+p_2^2-q_4^2}{2q_1p_2} \right)\;,
\eeq
where~$\mathcal P_\ell(x)$ are the Legendre polynomials. The volume of the shell then becomes
\begin{align}
V_s & = (2\pi)^3 \int_{p_i\in D_i} p_i^2 dp_i  \int_{q_j \in k_j} q_j^2 dq_j \sum_{\ell} (2\ell+1) \,  \nonumber \\
& \qquad \times \frac{1}{q_1q_4p_2} \mathcal P_\ell \left( \frac{q_1^2+p_2^2-q_4^2}{2q_1p_2} \right) \frac{1}{q_1q_2p_1} \mathcal P_\ell \left( \frac{q_1^2+q_2^2-p_1^2}{2q_1q_2} \right) \frac{1}{q_2q_3p_2} \mathcal P_\ell \left( \frac{q_2^2+p_2^2-q_3^2}{2q_2p_2} \right) \; .
\end{align}
Finally, the sum over $\ell$ can be done using 
\beq
\sum_\ell (2\ell+1) \mathcal{P}_\ell (\mu_1) \mathcal{P}_\ell (\mu_2) \mathcal{P}_\ell (\mu_3) = \frac{2}{\pi} \frac{1}{\sqrt{1 + 2\mu_1 \mu_2 \mu_3-\mu_1^2-\mu_2^2 -\mu_3^2}} \;.
\eeq
This equation has a nice geometric interpretation. It is related to the volume of the tetrahedron with three edges $q_1$, $q_2$ and $p_2$ 
\beq
V_{\rm tetra} = \frac{q_1 q_2 p_2}{6} {\sqrt{1 + 2\mu_1 \mu_2 \mu_3-\mu_1^2-\mu_2^2 -\mu_3^2}} \;,
\eeq
where $\mu_i$ are cosines of plane angles between three edges meeting in a common vertex. In our formula for $V_{1234}$ we have exactly the same form after closer inspections of all angles. Therefore, we can finally write (assuming that all bins are~$\Delta k$ wide)
\begin{align}\label{eq:shell}
V_s & = \frac83 \pi^2 \frac{k_1 k_2 k_3 k_4 D_1 D_2}{V_{\rm tetra}} \Delta k^6  \; .
\end{align}
The volume $V_{\rm tetra}$ can be computed in terms of six edges and this expression is manifestly symmetric in~$k_1,\ldots,k_4$ and~$D_1,D_2$.

Now we are ready to compute the trispectrum covariance matrix:

\begin{equation}
\begin{gathered}
\text{Cov}(k_1,k_2,k_3,k_4,D_1,D_2) = \langle\hat{T}(k_1,k_2,k_3,k_4,D_1,D_2)\hat{T}(k_a,k_b,k_c,k_d,D_a,D_b)\rangle-\\\langle\hat{T}(k_1,k_2,k_3,k_4,D_1,D_2)\rangle\langle\hat{T}(k_a,k_b,k_c,k_d,D_a,D_b)\rangle\\
=\frac{(2\pi)^{12}}{V^2 V_{1234} V_{abcd}} P(k_1)P(k_2)P(k_3)P(k_4)\prod_{i=1,a}^{4,d}\prod_{j=1,a}^{2,b} \int_{\bq_i\in\bk_i}d^3q_i\int_{\bp_j\in \mathbf{D}_j}d^3p_j\\
\times\delta_D(\bq_{1234})\delta_D(\bq_{12}+\p_1)\delta_D(\bq_{23}+\p_2)\delta_D(\bq_{abcd})\delta_D(\bq_{ab}+\p_a)\delta_D(\bq_{bc}+\p_b)\delta_D(\bq_{1a})\delta_D(\bq_{2b})\delta_D(\bq_{3c})\delta_D(\bq_{4d})+\text{23 perm.}\\
=\frac{(2\pi)^9}{V V_{1234} V_{abcd}} P(k_1)P(k_2)P(k_3)P(k_4)\delta^K_{\bk_1,\bk_a}\delta^K_{\bk_2,\bk_b}\delta^K_{\bk_3,\bk_c}\delta^K_{\bk_4,\bk_d}\delta^K_{\bp_1,\bp_a}\delta^K_{\bp_2,\bp_b}\prod_{i=a}^{d}\prod_{j=a}^{b} \int_{\bq_i\in\bk_i}d^3q_i\int_{\bp_j\in \mathbf{D}_j}d^3p_j\\
\times\delta_D(\bq_{abcd})\delta_D(\bq_{ab}+\p_a)\delta_D(\bq_{bc}+\p_b)+\text{23 perm.}\\
=\frac{(2\pi)^9}{V V_{1234}} P(k_1)P(k_2)P(k_3)P(k_4)\delta^K_{\bk_1,\bk_a}\delta^K_{\bk_2,\bk_b}\delta^K_{\bk_3,\bk_c}\delta^K_{\bk_4,\bk_d}\delta^K_{\bp_1,\bp_a}\delta^K_{\bp_2,\bp_b}+\text{23 perm.}\\
=\frac{(2\pi)^9R_{1234}}{V V_{1234}} P(k_1)P(k_2)P(k_3)P(k_4)\;,
\end{gathered}
\end{equation}
where $V_{1234}$ is given by Equation~\eqref{eq:shell}.
\subsection{PT cross terms}\label{app:PTterms}
\noindent In the Gaussian approximation some PT cross-terms survive.
We have:
\begin{equation}
    \begin{gathered}
        \langle T(k_1,k_2,k_3,k_4,D_1,D_2)P(k)\rangle - \langle T(k_1,k_2,k_3,k_4,D_1,D_2)\rangle \langle P(k)\rangle=\\
        \frac{1}{V^2V_{1234}V_P}\prod_{i=1}^{4}\prod_{j=1,2}\prod_{n=a,b} \int_{q_i\in k_i}{d^3q_i}\int_{p_i\in D_j}{d^3p_j}\int_{q_n\in k_n}{d^3q_n}\delta_D(\bq_{1234}) \delta_D(\bq_{ab}) \delta_D(\bq_{12}+\bp_1) \delta_D(\bq_{23}+\bp_2)\times \\
        \langle(\delta(\bq_1) \delta(\bq_2) \delta(\bq_3) \delta(\bq_4) \delta(\bq_a) \delta(\bq_b)\rangle_c   \\
        =2\frac{(2\pi)^9}{V^2V_{1234}V_P}\prod_{i=1}^{4}\prod_{j=1,2}\prod_{n=a,b} \int_{q_i\in k_i}{d^3q_i}\int_{p_i\in D_j}{d^3p_j}\int_{q_n\in k_n}{d^3q_n}\delta_D(\bq_{1234}) \delta_D(\bq_{ab}) \delta_D(\bq_{12}+\bp_1) \delta_D(\bq_{23}+\bp_2)\times \\
       \delta_D(\bq_{1a}) \delta_D(\bq_{3b})\delta_D(\bq_{24})P(q_1)P(q_3)P(q_2)\\
       = 2\frac{(2\pi)^6}{VV_{1234}V_P}P^2(k)P(k_2)\delta_{\bk_,\bk_1}\delta_{\bk_,\bk_3}\delta_{\bk_2,\bk_4}\prod_{i=3,4}\prod_{j=1,2} \int_{q_i\in k_i}{d^3q_i}\int_{p_i\in D_j}{d^3p_j}\delta_D(-\bq_3-\bq_4+\bp_1)\delta_D(\bq_3-\bq_4+\bp_2)\;,      
    \end{gathered}
\end{equation}
all the other contractions will give $\bp_{1,2}=0$ which do not fall in any bin. Now we want to evaluate:
\beq
\begin{gathered}
    \prod_{i=3,4}\prod_{j=1,2} \int_{q_i\in k_i}{d^3q_i}\int_{p_i\in D_j}{d^3p_j}\delta_D(-\bq_3-\bq_4+\bp_1)\delta_D(\bq_3-\bq_4+\bp_2)\\
    =\frac{1}{(2\pi^2)^2} \int d^3q_3\int d^3q_4\int dp_1p_1^2\int dp_2p_2^2\int d^3 x\int d^3y e^{i\bq_3(\bx-\mathbf{y})}e^{-i\bq_4(\bx+\mathbf{y})}j_0(p_1y)j_0(p_2x)\;,
\end{gathered}
\eeq
where we performed angular integrations over $\bp_1$ and $\bp_2$. Using Equation~\eqref{eq:exp} leads to
\beq
\begin{gathered}
    2^6\sum_{\ell}(2\ell+1)\int dq_3 dq_4 dp_1dp_2 dxdyq_3^2q_4^2p_1^2p_2^2x^2y^2 j_{\ell}(q_3x)j_{\ell}(q_3y)j_{\ell}(q_4x)j_{\ell}(q_4y)j_0(p_1x)j_0(p_2y)\;,
\end{gathered}
\eeq

and using Equation~\eqref{eq:legendre} we get:
\beq\label{eq:eq}
\begin{gathered}
    4\pi^2\sum_{\ell}(2\ell+1)\int p_1p_2 dq_3 dq_4 dp_1dp_2\mathcal{P}_{\ell}\left(\frac{q_3^2+q_4^2-p_1^2}{2q_3q_4}\right)\mathcal{P}_{\ell}\left(\frac{q_3^2+q_4^2-p_2^2}{2q_3q_4}\right)\;.
\end{gathered}
\eeq
Now it is known that
\beq
\begin{gathered}
\sum_{\ell}\frac{(2\ell+1)}{2}\mathcal{P}_{\ell}(\mu_1)\mathcal{P}_{\ell}(\mu_2)=\delta_D(\mu_1-\mu_2)\;.
\end{gathered}
\eeq

So Equation~\eqref{eq:eq} reads
\beq
\begin{gathered}
    2(4\pi^2)\delta_{D_1,D_2}\int p_1^2dp_1\int dq_3\int dq_4= 8 \pi^2\delta_{D_1,D_2}D_1^2\Delta k^3\;.
\end{gathered}
\eeq
So PT terms are non-zero only for quadrilaterals which have $k_1=k_3$,  $k_2=k_4$, $D_1=D_2$; given that we are considering quadrilaterals with ordered sides, the only possible configurations are the equilateral shapes, i.e. $k_1 = k_2= k_3 = k_4$ and $D_1 = D_2$. They are $\sim 0.1\%$ of the total number.

\bibliography{ref}

\end{document}